\documentclass[twocolumn]{aastex631}
\usepackage{graphics, graphicx}
\usepackage{subfigure}
\usepackage{threeparttable}
\usepackage{multirow}
\usepackage{amsthm, amsmath}
\usepackage[mathlines]{lineno}
\usepackage[normalem]{ulem}
\usepackage{soul}

\def\cxl#1 {{\textcolor{blue}{#1}}\ }
\newcommand{\dmhost}{DM$_{\rm host}$}
\newcommand{\dmfrb}{DM$_{\rm FRB}$}
\newcommand{\dmcosmic}{DM$_{\rm IGM}$}
\newcommand{\dmmw}{DM$_{\rm MW}$}
\newcommand{\dmhalo}{DM$_{\rm halo}$}
\newcommand{\dmdisk}{DM$_{\rm disk}$}

\def\OIII{O\,{\sc iii}}

\def\Ha{H{$\rm{\alpha}$}}
\def\Hb{H{$\rm{\beta}$}}

\def\NII{N\,{\sc ii}}
\def\SII{S\,{\sc ii}}

\shorttitle{Host Galaxy of FRB 20240114A}
\shortauthors{Chen et al.}
\graphicspath{{./}{figures/}}

\begin{document}

\title{The Host Galaxy of the Hyper-Active Repeating FRB 20240114A: Behind a Galaxy Cluster}

\correspondingauthor{C.-W. Tsai; D. Li}
\email{cwtsai@nao.cas.cn; dili@tsinghua.edu.cn}

\author[0000-0001-5738-9625]{Xiang-Lei Chen}
\affiliation{National Astronomical Observatories, Chinese Academy of Sciences, 20A Datun Road, Beijing 100101, China}
\affiliation{Key Laboratory of Radio Astronomy and Technolgoy, Chinese Academy of Sciences, A20 Datun Road, Chaoyang District, Beijing, 100101, P.R. China}

\author[0000-0002-9390-9672]{Chao-Wei Tsai}
\affiliation{National Astronomical Observatories, Chinese Academy of Sciences, 20A Datun Road, Beijing 100101, China}
\affiliation{Institute for Frontiers in Astronomy and Astrophysics, Beijing Normal University,  Beijing 102206, China}
\affiliation{University of Chinese Academy of Sciences, Beijing 100049, China}

\author[0000-0003-3010-7661]{Di Li}
\affiliation{Department of Astronomy, Tsinghua University, Beijing 100084, China}
\affiliation{National Astronomical Observatories, Chinese Academy of Sciences, 20A Datun Road, Beijing 100101, China}

\author[0000-0002-3386-7159]{Pei Wang}
\affiliation{National Astronomical Observatories, Chinese Academy of Sciences, 20A Datun Road, Beijing 100101, China}
\affiliation{Institute for Frontiers in Astronomy and Astrophysics, Beijing Normal University,  Beijing 102206, China}

\author[0000-0002-0475-7479]{Yi Feng}
\affiliation{Research Center for Astronomical Computing, Zhejiang Laboratory, Hangzhou 311100, China}
\affil{Institute for Astronomy, School of Physics, Zhejiang University, Hangzhou 310027, China}

\author[0009-0005-8586-3001]{Junshuo Zhang}
\affiliation{National Astronomical Observatories, Chinese Academy of Sciences, 20A Datun Road, Beijing 100101, China}
\affiliation{University of Chinese Academy of Sciences, Beijing 100049, China}

\author[0000-0003-4007-5771]{Guodong Li}
\affiliation{National Astronomical Observatories, Chinese Academy of Sciences, 20A Datun Road, Beijing 100101, China}
\affiliation{University of Chinese Academy of Sciences, Beijing 100049, China}

\author[0000-0002-8744-3546]{Yongkun Zhang}
\affiliation{National Astronomical Observatories, Chinese Academy of Sciences, 20A Datun Road, Beijing 100101, China}

\author[0009-0009-6319-0888]{Lulu Bao}
\affiliation{University of Chinese Academy of Sciences, Beijing 100049, China}

\author[0000-0002-9137-7019]{Mai Liao}
\affiliation{National Astronomical Observatories, Chinese Academy of Sciences, 20A Datun Road, Beijing 100101, China}
\affiliation{Universidad Diego Portales, Av Republica 180, Santiago, Reg\'{i}\'{o}n Metropolitana, Chile}

\author[0009-0002-3093-0180]{Ludan Zhang}
\affiliation{National Astronomical Observatories, Chinese Academy of Sciences, 20A Datun Road, Beijing 100101, China}
\affiliation{University of Chinese Academy of Sciences, Beijing 100049, China}

\author[0000-0003-3948-9192]{Pei Zuo}
\affiliation{National Astronomical Observatories, Chinese Academy of Sciences, 20A Datun Road, Beijing 100101, China}

\author[0000-0003-2024-1648]{Dongwei Bao}
\affiliation{National Astronomical Observatories, Chinese Academy of Sciences, 20A Datun Road, Beijing 100101, China}

\author[0000-0001-6651-7799]{Chenhui Niu}
\affiliation{Institute of Astrophysics, Central China Normal University, Wuhan 430079, China}

\author[0000-0002-4300-121X]{Rui Luo}
\affiliation{Department of Astronomy, School of Physics and Materials Science, Guangzhou University, Guangzhou 510006, China}

\author[0000-0001-5105-4058]{Weiwei Zhu}
\affiliation{National Astronomical Observatories, Chinese Academy of Sciences, 20A Datun Road, Beijing 100101, China}
\affiliation{University of Chinese Academy of Sciences, Beijing 100049, China}

\author[0000-0002-6684-3997]{Hu Zou}
\affiliation{National Astronomical Observatories, Chinese Academy of Sciences, 20A Datun Road, Beijing 100101, China}

\author[0000-0002-0660-0432]{Suijian Xue}
\affiliation{National Astronomical Observatories, Chinese Academy of Sciences, 20A Datun Road, Beijing 100101, China}

\author[0000-0002-9725-2524]{Bing Zhang}
\affiliation{Nevada Center for Astrophysics, University of Nevada, Las Vegas, NV 89154, USA}
\affiliation{Department of Physics and Astronomy, University of Nevada, Las Vegas, NV 89154, USA}

\begin{abstract}

We report on the optical spectroscopic observations of the host galaxy of the hyperactive repeating fast radio burst, FRB 20240114A. 
The host galaxy is a dwarf galaxy at a redshift of $z=0.1306\pm0.0002$. With a rest-frame coverage of 4300\AA--7900\AA, we have detected \Ha, \Hb, [\OIII]\,$\lambda\lambda$4959,5007, [\NII]$\lambda\lambda$6548,6583, and [\SII]\,$\lambda$6716 emission lines. The emission line ratios suggest that the ionization in the host galaxy is dominated by star formation. 
The star formation rate (SFR) derived from the \Ha\ emission line is $(0.06 \pm 0.01) \ \rm{M_{\odot} \ yr^{-1}}$, and the SED fitting suggests the lower limit of the SFR(UV) is $0.09 \ \rm{M_{\odot} \ yr^{-1}}$. 
The stellar mass is $(\rm 4.0 \pm 1.8) \times 10^8 \ M_{\odot}$, making the specific star formation rate $\rm log \ sSFR(H\rm \alpha) = -9.8 \pm 0.2 \ yr^{-1}$. 
The line ratios indicate an upper limit of a metallicity of $\rm 12+log_{10} ([O/H]) \sim 8.5$.
As the nearest dwarf host galaxy with a repeating FRB, the activity of FRB 20240114A and the properties of this host galaxy closely resemble those of FRB 20121102A and FRB 20190520B. 
The \Ha -traced dispersion measure (DM) provided by the ionized gas of the host galaxy has a moderate contribution of $\sim 200 \rm \ pc \ cm^{-3}$, assuming a warm ionized gas. 
We found that the distributions of the stellar mass versus SFR are significantly different between repeating and one-off FRBs, as determined by the MANOVA test with $p=0.0116$.

\end{abstract}


\section{Introduction} 
\label{sec:intro}

FRB 20240114A, a hyperactive repeating fast radio burst (FRB), was recently identified by CHIME \citep{CHIMEdiscovery2024ATel16420....1S}. 
Three bursts were observed within 10 days, each with an exposure of 4 minutes, in the 400--800 MHz band of CHIME.
The brightest burst exhibited a fluence reaching 919 $\pm$ 97 Jy ms (1$-\sigma$ uncertainty), averaged over the entire CHIME band. 
The Five-hundred-meter Aperture Spherical radio Telescope (FAST) monitoring of FRB 20240114A reached an average burst rate of approximately 500 $\rm h^{-1}$ during a burst storm between MJD 60374 and MJD 60386, with fluences exceeding the 0.015 Jy ms threshold of FAST (Zhang et al., in prep).
These suggest that FRB 20240114A has the potential to become a highly active repeating FRB with strong fluence.

\begin{figure}[ht]
    \centering
    \includegraphics[width=0.95\linewidth]{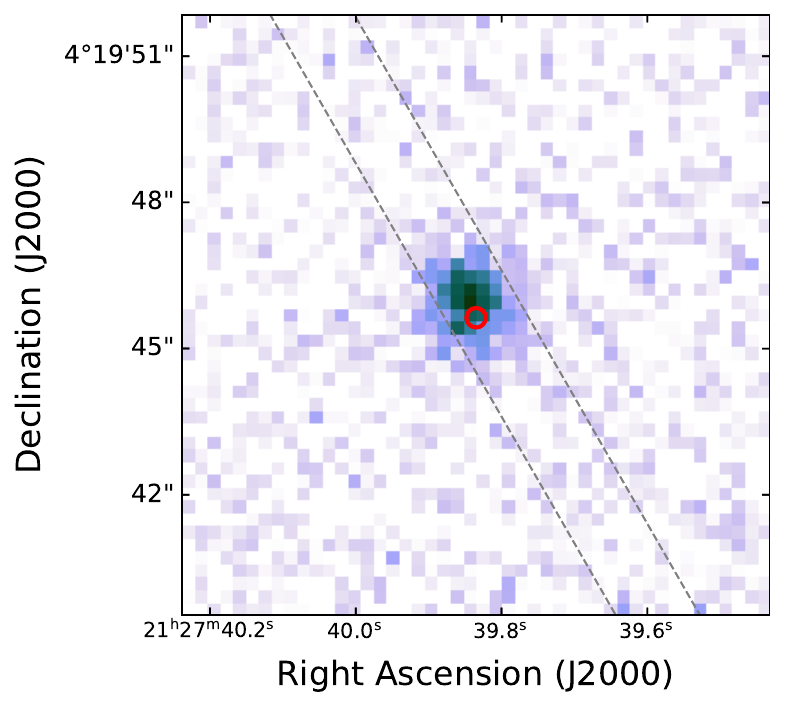}
    \caption{Legacy Survey $r$-band image \citep[DR9,][]{LS2019AJ....157..168D} of the host galaxy of FRB 20240114A.
    The red circle marks the FRB location with an uncertainty of $r=0.2^{\prime\prime}$. The dashed gray lines depict the slit configuration used in the spectroscopic observation.}
    \label{fig:hostimg}
\end{figure}

The dispersion measure (DM) of FRB 20240114A ranged from 527.1 to 531.3 pc cm$^{-3}$. The reported rotation measure (RM) of the source varied between 325 and 360 rad m$^2$ \citep{2024ATel16426....1O, 2024ATel16430....1U, 2024ATel16432....1O, 2024ATel16433....1Z, 2024ATel16434....1P, 2024ATel16446....1T, 2024ATel16452....1K, 2024ATel16494....1P, 2024ATel16505....1Z}. 
After accounting for the Milky Way's contribution to the DM, the remaining extragalactic DM translates to a redshift upper limit of $z \sim 0.5$ \citep{MacquartRelation2020Natur.581..391M}.

The localization of FRB 20240114A was carried out by MeerKAT \citep{2024ATel16446....1T} and the European VLBI Network \citep[EVN;][]{2024ATel16542....1S}, in that order.
The precise coordinates of the FRB, as determined by the EVN, are (R.A., decl.) [J2000] = (21:27:39.835, +04:19:45.634) with an uncertainty of 0.2$^{\prime\prime}$ \citep{2024ATel16542....1S}.
The FRB was spatially coincident with the galaxy SDSS J212739.84$+$041945.8 \citep{SDSS2015ApJS..219...12A}, exhibiting a Probabilistic Association of Transients to their Hosts \cite[PATH;][]{PATH2021ApJ...911...95A} probability of 0.9974 \citep{Tian2024MNRAS.533.3174T}. 
Figure~\ref{fig:hostimg} displays the $r$-band image of the host galaxy, with the FRB localized in the southwestern region of the galaxy.
The photometric redshift (photo-$z$) of the host galaxy is $0.420 \pm 0.1568 $ according to SDSS and $0.252 \pm 0.124$ according to Legacy Survey DR9. 
Both redshift estimations fall within the redshift upper limit derived from the extragalactic DM.
The spectroscopic redshift, $z = 0.1300 \pm 0.0002$, was reported in \cite{2024ATel16613....1B}. 
Utilizing MeerKAT, the discovery of a radio continuum coincident with the location of FRB 20240114A identifies it as a new candidate for a repeating FRB associated with a persistent radio source \citep[PRS;][]{2024ATel16695....1Z}.
Subsequent observations further support the identification of PRS candidates \citep{Bruni2024arXiv241201478B, uGMRT2024arXiv241213121B, PRSYu2025arXiv250114247Z}.

The low-mass property of the host galaxy was inferred before the spectroscopic redshift was determined.
Leveraging photometric data from the Legacy Survey and considering the redshift upper limit of 0.5, the stellar mass of the host galaxy of FRB 20240114A is estimated to be as low as $10^{9} \ M_{\bigodot}$ \citep{Kauffmann2003MNRAS.341...33K}.
This low stellar mass makes the host galaxy a potential candidate for the fourth dwarf FRB host galaxy, following the hosts of FRB 20121102A \citep{Spitler2014ApJ...790..101S, Spitler2016Natur.531..202S, Chatterjee121102-2017Natur.541...58C, Tendulkar121102-2017ApJ...834L...7T}, FRB 20190520B \citep[][Chen et al. submitted to ApJ]{Niu190520-2022Natur.606..873N}, and FRB 20210117A \citep{Bhandari2101172022arXiv221116790B}.
It is noteworthy that both FRB 20121102A and FRB 20190520B are highly active repeating FRBs associated with a luminous PRS, with a spectral luminosity of $ L_{\nu} > 10^{29} \ \rm erg \ s^{-1} \ Hz^{-1} $, which is not attributed to star formation in their host galaxies \citep{PRS2022ApJ...927...55L}.
(We also note that FRB 20181030A and FRB 20190417A were recently discovered to be associated with PRS candidates, \citealp{newPRS2024arXiv240911533I}.)
The host galaxies of FRB 20121102A and FRB 20190520B have been identified as low-metallicity galaxies.
\cite{Tendulkar121102-2017ApJ...834L...7T} used emission line ratios to estimate the metallicity of the host galaxy of FRB 20121102A to be $\rm 12 + log_{10} ([O/H]) \leq 8.7$ on the KK04 scale \citep{KK042004ApJ...617..240K}, which indicates that only $\sim$15\% of galaxies brighter than $M_B < -16$ have metallicities lower than 8.7 \citep{Graham2017ApJ...834..170G}.
The metallicity of the host galaxy of FRB 20190520B is estimated to be $\rm 12 + log_{10} ([O/H]) \leq 7.4 \pm 0.4$ (Chen et al. submitted to ApJ), derived using the $T_e$ method \citep{Aller1984ASSL..112.....A, Izotov2006A&A...448..955I}.
FRB 20210117A, on the other hand, is currently classified as an one-off FRB, exhibiting a unique downward-drifting structure in its dynamic spectrum, a characteristic now recognized as a signature of repeating FRBs \citep{Pleunis2021ApJ...923....1P}.
The observed similarities among FRB 20240114A, FRB 20121102A and FRB 20190520B in terms of high activity, and the similar characteristics of the dwarf host galaxies, may provide valuable insights into understanding the FRB evolutionary state and the stellar populations responsible for their progenitors.

This paper presents the optical spectroscopic observations and properties of the host galaxy of FRB 20240114A.
The observations and the data reduction results are presented in Sect.~\ref{sect:obs}. 
Sect.~\ref{sect:gal-ana} analyzes the host galaxy properties . 
Sect.~\ref{sect:dissc} discusses the galaxy clustering and the progenitor of FRB 20240114A. 
Sect.~\ref{sect:summary} provides a summary of the findings.
The data of the host galaxies of FRBs used in this work is up to May 2024.
We use the flat $\Lambda$CDM model with $H_0 = 67.66$ km s$^{-1}$ Mpc$^{-1}$ and $\Omega_{\rm m}=0.310$ \citep{Planck2020A&A...641A...6P}. 
Solar abundance of $\rm 12+log_{10} ([O/H]) = 8.69 \pm 0.04$ is adopted \citep{solar3-2021A&A...653A.141A}.

\section{Observations}
\label{sect:obs}

\subsection{GTC Spectroscopic Observation}
\label{subsect:spec_obs}

We obtained optical spectroscopic observations with the Gran Telescopio CANARIAS (GTC) under the target of opportunity (ToO) program GTCMULTIPLE3A-24ACNT (PI: C-W Tsai). 
We activated the trigger on March 15th, 2024, and the observations were executed on May 3rd, 2024. 
The observation used a long slit with a width of 1.5$^{\prime \prime}$ and the OSIRIS+/R500R grism covering 4800$-$10000 \AA. 
The default $2 \times 2$ binning is applied. 
The position angle for the slit was set at 30 degrees from North to East along the elongated direction of the galaxy in the Legacy Survey $r$-band image. 
We obtained a in total of 3000 s of integration on the source, including 3 frames with an offset of 10$^{\prime \prime}$, under good weather conditions (seeing$\sim 0.9^{\prime \prime}$.) 
The standard star Feige 66 was used for the flux calibration. 
The wavelength calibration was performed with the arc lines of HgAr, Ar and Ne lamps.

\begin{figure*}[ht!]
    \centering
    \includegraphics[width=0.95\textwidth]{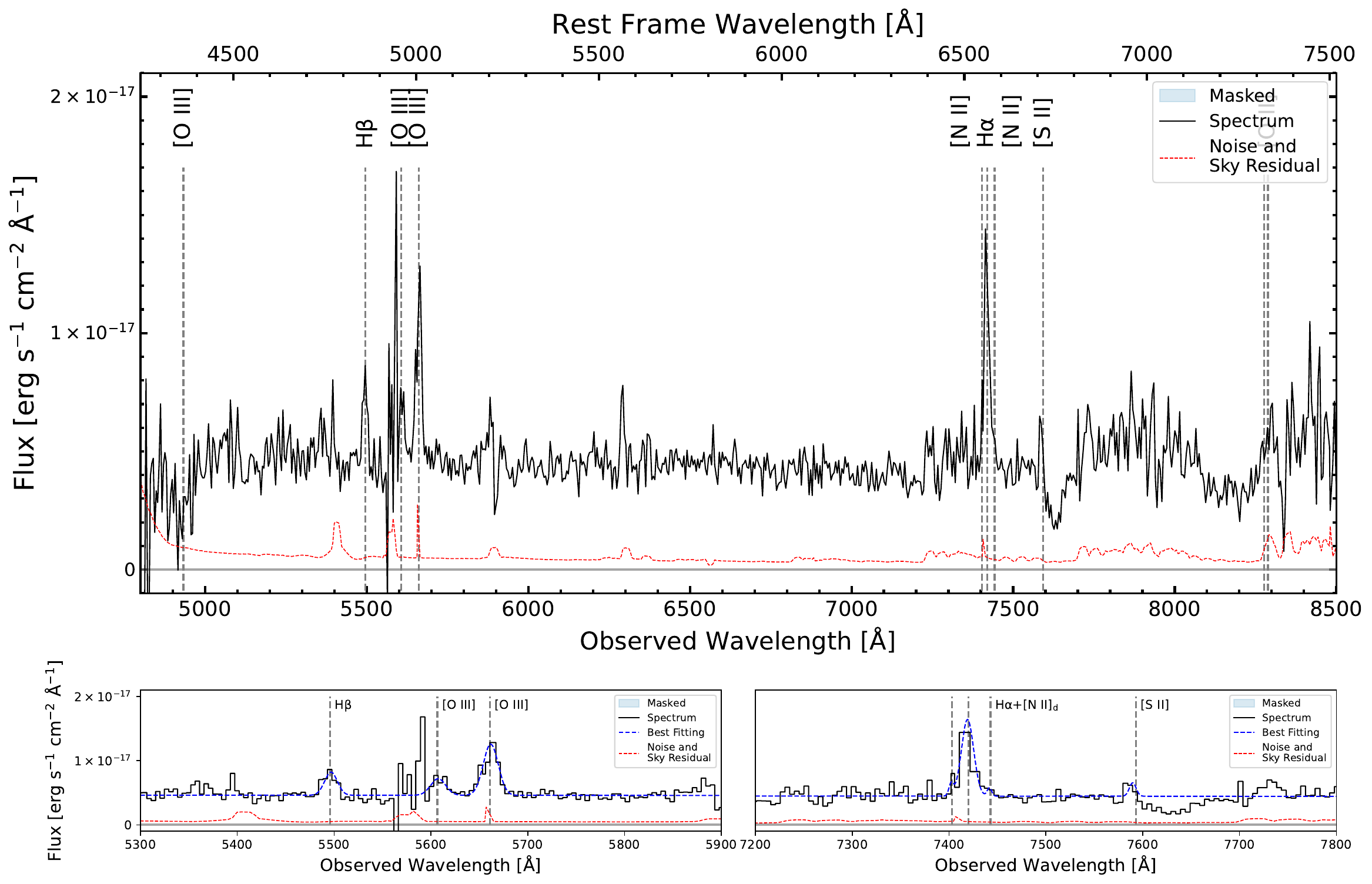}
    \caption{
    GTC/OSIRIS+ R500R spectrum of the FRB 20240114A host galaxy at $z = 0.1306$.
    The noise, including the sky residual, is represented by the red dashed line, and the blue shaded regions represent the masked areas in the emission line fitting due to noise and telluric absorption around 7600 \AA.
    The lower panels show the best fitting results of the emission lines of \Hb, [\OIII]$\lambda\lambda$4959,5007 (left) and \Ha, [\NII]$\lambda\lambda$6548,6583, [\SII]$\lambda$6716 (right), respectively.
    The blue dashed lines represent the best fittings. 
    }
    \label{fig:gtc-spec}
\end{figure*}

We use PyPeit \citep{pypeit1, pypeit2} to reduce the spectroscopic data using global sky subtraction \footnote{\url{https://pypeit.readthedocs.io/en/release/skysub.html}}. 
Figure~\ref{fig:gtc-spec} shows the co-added spectrum of all three exposures.
The spectrum of the host galaxy of FRB 20240114A shows emission lines of \Ha, \Hb, [\OIII]$\lambda\lambda$4959,5007, [\NII]$\lambda\lambda$6548,6583, [\SII]$\lambda$6716.
We fit the emission lines using single-Gaussian profiles and the continuum using a power law to describe the galaxy contribution. 
Due to the noise around 5600 \AA\ and the telluric absorption around 7600 \AA, the line profiles of [\OIII] $\lambda$4959 and [\SII] $\lambda$6716 may not be directly fitted. 
Therefore, we fix the line width of [\OIII] $\lambda$4959 to be the same as that of [\OIII] $\lambda$5007, and fit the [\SII] $\lambda$ 6716 line to provide an upper limit. 
The fitting of the [\NII]$\lambda\lambda$6548,6583 lines is affected by the \Ha, so we fit the [\NII] $\lambda\lambda$6548,6583 together with the same width. 
Blue shadows in Figure~\ref{fig:gtc-spec} represent the masked noisy regions during the fitting, and Table~\ref{tab:em-lines} shows the fitting results of the emission lines. 
The fitted redshift of the host galaxy is $z=0.1306 \pm 0.0002$, and is consistent with the previous report in \cite{2024ATel16613....1B}.

\begin{table}[ht]
\centering
    \caption{Emission line fluxes of the host galaxy of FRB 20240114A. The $1-\sigma$ uncertainties are in parentheses.
    }
    \medskip
    \begin{tabular*}{0.9\linewidth}{lccc}
        \hline
        \hline
        \noalign{\smallskip}
        Line                      & FWHM ($ \sigma $)      & F$ _{ \lambda } $ ($ \sigma $)  \\
                                  & (\AA)                  & ($ 10^{-17} $ erg cm$ ^{-2} $ s$ ^{-1} $)  \\
        \noalign{\smallskip}
        \hline
        \noalign{\smallskip}
        \Ha                & 14.96 (1.21)            & 25.05 (2.70) \\
        \noalign{\smallskip}
        \Hb                 & 15.52 (3.64)            & 8.56 (1.72)  \\
        \noalign{\smallskip}
        [\OIII] $ \lambda $ 4959   & \multirow{2}{*}{18.31 (1.72)} &  7.45 (1.64) \\
        \noalign{\smallskip}
        [\OIII] $ \lambda $ 5007   &                        & 22.77 (2.94) \\
        \noalign{\smallskip}
        [\NII] \ $ \lambda $ 6548  & \multirow{2}{*}{5.37 (3.60)} & 1.46 (1.32) \\ 
        \noalign{\smallskip}
        [\NII] \ $ \lambda $ 6583  &                        & 0.96 (0.96) \\
        \noalign{\smallskip}
        \hline
    \end{tabular*}
    \label{tab:em-lines}
\end{table}

\subsection{Multi-band Photometry}
\label{subsect:photometry}
 
In order to construct the spectral energy distribution (SED), we assembled the host galaxy's multi-wavelength broad-band photometry spanning ultraviolet (UV) to optical wavelengths. 
We cross-matched it with the \textit{GALEX} Medium-deep Sky Catalog \citep{2017ApJS..230...24B} and found an object at a distance of $1.7^{\prime\prime}$, with an FUV ($1350-1780$ \AA) flux of $(1.4\pm0.5)\ \mu$Jy and an NUV ($1770-2730$ \AA) flux of $(1.1\pm0.4)\ \mu$Jy. 
To perform a more reliable SED fitting, we use the reduced and calibrated photometric data from the \textit{GALEX} archive\footnote{All the \textit{GALEX} data used in this paper can be found in MAST: \dataset[10.17909/dt3d-av21]{http://dx.doi.org/10.17909/dt3d-av21}.} to re-measure the FUV and NUV flux. 
A $2.0^{\prime\prime}$ diameter aperture was placed on the target, centered on the position R.A. $= 21^{\rm h}27^{\rm m}39^{\rm s}.84$, decl. $= +04^{\circ}19^{\prime}45.8^{\prime\prime}$ (J2000).
We obtained an FUV flux of $(0.8\pm0.3)\ \mu$Jy and an NUV flux of $(0.8\pm0.4)\ \mu$Jy, with the FUV flux showing a significant difference from the \textit{GALEX} pipeline result. 
We checked the calibration files and found no anomalies with the instrument or observations. 
The deep optical images from the Legacy Survey do not show any structure $\rm 1.7^{\prime\prime}$ to the northwest of the center. 
Therefore, we adopt the manually measured results for the SED fitting in this paper.

For the optical photometry, we use the released data of the Sloan Digital Sky Survey \citep[DR18;][]{2023ApJS..267...44A}, the Pan-STARRS Survey \citep[DR2;][]{2016arXiv161205560C}, and the Legacy Survey \citep[DR10;][]{LS2019AJ....157..168D}. 
For data from the same band but from the different optical surveys, we adopt the values with lower uncertainties among the surveys.
The flux densities of the multi-wavelength photometric data are given in Table~\ref{tab:table_1}.
We note that the host galaxy of FRB 20240114A is not detected in the AllWISE or CatWISE2020 catalog \citep{2020ApJS..247...69E,2021ApJS..253....8M}, although it is covered by the \textit{Wide-field Infrared Survey Explorer (WISE)} survey \citep{2010AJ....140.1868W}. 
The rest-frame luminosities derived from the upper limits of the WISE images are several orders of magnitude higher than those from the other bands.
Therefore, we did not use these limits in the SED fitting due to their poor constraints.
The non-detection in the infrared bands may indicate a low dust extinction of the host galaxy.

\begin{table}[htbp]
\centering
    \caption{Flux Densities of the host galaxy of FRB 202401114A.
    }
    \medskip
    \begin{tabular*}{0.9\linewidth}{lccc}
        \hline
        \hline
        \noalign{\smallskip}
        Band                      &Coverage & $\lambda_{\rm eff.}$ & Flux Density  \\
                                  & (\micron)       & (\micron)      & ($\rm{\mu Jy}$)  \\
        \noalign{\smallskip}
        \hline
        \noalign{\smallskip}
        GALEX.fnu.                & 0.135$-$0.178 & 0.15.      & 0.8$\pm$0.3 \\
        \noalign{\smallskip}
        GALEX.nuv                 &	0.177$-$0.273 & 0.23 	& 0.8$\pm$0.4  \\
        \noalign{\smallskip}
        SDSS.u.                   & 0.305$-$0.403 & 0.36       & 2.2$\pm$0.9 \\
        \noalign{\smallskip}
        Legacy.g 	              & 0.385$-$0.557 & 0.48	& 4.2$\pm$0.1 \\
        \noalign{\smallskip}
        Legacy.r 	              & 0.540$-$0.730 & 0.62 	& 6.7$\pm$0.1 \\ 
        \noalign{\smallskip}
        PS1.i 	                  & 0.678$-$0.831 & 0.75 	& 7.5$\pm$0.6 \\ 
        \noalign{\smallskip}
        Legacy.z 	              & 0.825$-$1.015 & 0.87 	& 8.3$\pm$0.4 \\
        \noalign{\smallskip}
        \hline
    \end{tabular*}
    \label{tab:table_1}
\end{table}

We examined the major data archives, including ALMA, Gemini, HST, JWST, Keck, Subaru, VLA, and VLT (from radio to optical), and we did not find any other detections within a radius of 20$^{\prime\prime}$ centered on FRB 20240114A listed before August 2024. 
Moreover, no detections were found within a 6$^{\prime}$ radius of FRB 20240114A in the catalogs from \emph{Chandra}, eROSITA and XMM. 
Additionally, a search of the Meta-Catalog of the compiled properties of X-ray detected Clusters of galaxies (MCXC), which is based on the publicly available ROSAT All Sky Survey \citep{MCXC2011A&A...534A.109P}, revealed no galaxy cluster within the same radius.

\section{Properties of the Host Galaxy}
\label{sect:gal-ana}

\subsection{Spectroscopic Analysis}
\label{subsect:spectrum}

\begin{figure}[ht]
    \centering
    \includegraphics[width=0.95\linewidth]{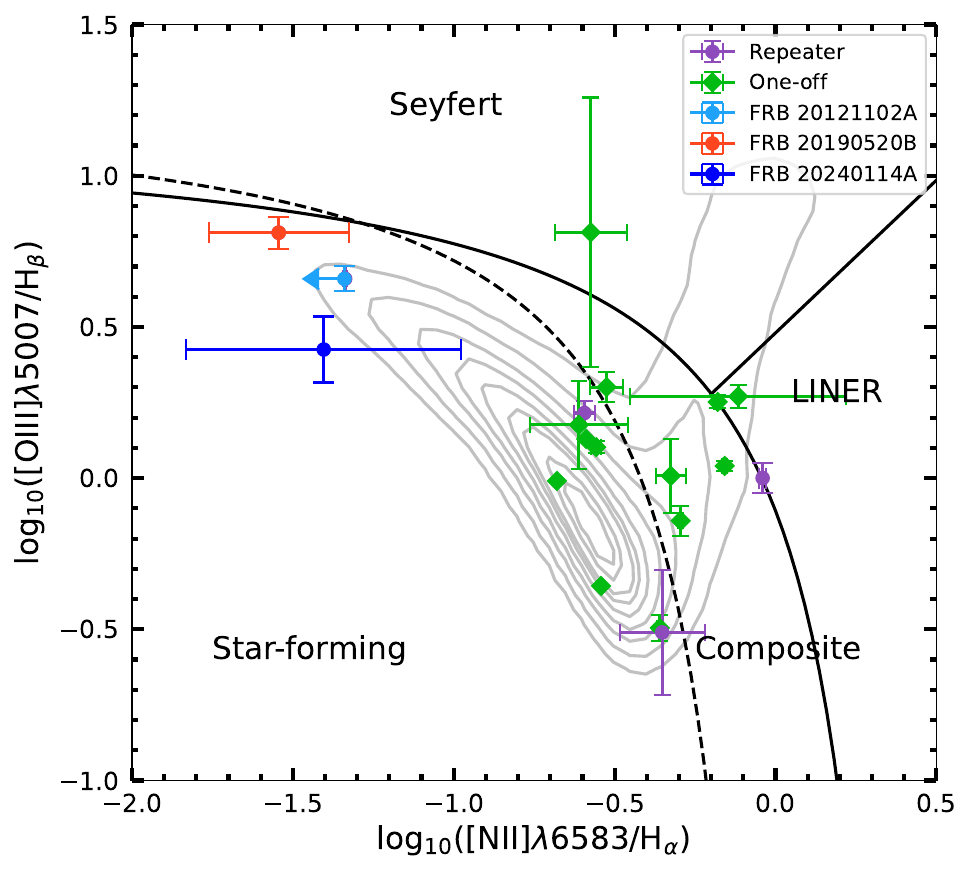}
    \caption{BPT classification diagram
    for the host galaxy of FRB 20240114A and that of other FRBs.
    The contours represent the SDSS DR17 galaxies with significant emission lines \citep[$>5\sigma$,][]{SDSSDR172022ApJS..259...35A}. The lines indicate the demarcation between star-forming galaxies and composite sources \citep{Kauffmann2003MNRAS.346.1055K}, composite sources and LINERs \citep{Kewley2006MNRAS.372..961K}, and LINER and Seyfert galaxies \citep{Fernandes2010MNRAS.403.1036C}, respectively. }
    \label{fig:bpt}
\end{figure}

Figure~\ref{fig:bpt} shows the Baldwin, Phillips \& Terlvich (BPT) diagram \citep{BPT1981PASP...93....5B} for the host galaxy of FRB 20240114A derived using the fluxes presented in Table~\ref{tab:em-lines}, together with other FRB host galaxies. 
The host galaxy of FRB 20240114A is classified as a star-forming galaxy and resides in proximity to the dwarf host galaxies of other repeating FRBs, namely FRB 20121102A and FRB 20190520B. 
The star formation rate estimated using the \Ha\ emission line is $0.06 \pm 0.01 \ \rm{M_{\odot}} \ \rm yr^{-1}$ \citep{2012ARA&A..50..531K}.
We estimate the Galactic extinction along this line of sight to be $E_{(B-V)}=0.058$.
Using the line flux ratio between \Ha\ and \Hb, we estimate the intrinsic extinction to be $A_V = 0.220$, assuming a case-B line ratio of 2.86.
The low intrinsic extinction of the host galaxy of FRB 20240114A is commonly seen in low-mass star-forming galaxies \citep[e.g.][]{Griffith2011ApJ...736L..22G}. 

The metallicity of the host galaxy of FRB 20240114A is estimated using the diagnostics from PP04 \citep{Alloin1979, PP042004MNRAS.348L..59P}
\begin{equation}
    N2 \equiv \rm log_{10}([N\,{\textsc{ii}}]\lambda 6583/ \rm H\alpha)
\end{equation}
and
\begin{equation}
    O3N2\equiv \rm log_{10} (\frac{[O\,{\textsc{iii}}]\lambda 5007) / \rm H\beta}{[N\,{\textsc{ii}}]\lambda 6583 / \rm H\alpha}),
\end{equation}
since the emission line of [\OIII]$\lambda$4363 required to estimate the metallicity using the T$_e$ method \citep{Aller1984ASSL..112.....A, Izotov2006A&A...448..955I} is outside the spectroscopic wavelength coverage,
The averaged metallicity is $12 + \rm log_{10}([O/H]) \sim 8.5 $ in KK04 scale \citep{KK042004ApJ...617..240K}.
This metallicity is lower than that of the host galaxy of FRB 20121102A with $\rm 12 + log_{10} ([O/H]) \leq 8.7$ applying a similar estimation and the same scale \citep{Tendulkar121102-2017ApJ...834L...7T}.

\subsection{Ionized Gas Contribution to DM} 
\label{subsect:halphaDM}

The observed dispersion measure of FRB 20240114A \dmfrb\ can be expressed by 
\begin{equation}
    \rm DM_{FRB} = DM_{MW} + DM_{IGM} + DM_{host},
\end{equation}

where \dmmw, $\rm DM_{IGM}$ and \dmhost\ are the contributions on the line of sight from Milky Way, intergalactic medium (IGM) and FRB host galaxy, respectively \citep{ZhangDMz2018ApJ...867L..21Z}.
The \dmmw\ consists of contributions from Milky Way halo \citep[\dmhalo, $\sim$45 pc cm$^{-3}$,][]{DMhalo2020ApJ...888..105Y} and disk (\dmdisk).
The \dmdisk\ is estimated as 49.7 pc cm$^{-3}$ \citep[NE2001,][]{NE2001-12002astro.ph..7156C, NE2001-22003astro.ph..1598C} or 38.8 pc cm$^{-3}$ \citep[YMW16,][]{YMW2017ApJ...835...29Y} according to different models. 
To estimate \dmcosmic, we apply the averaged relationship in \cite{ZhangDMz2018ApJ...867L..21Z}
\begin{equation}
    z \sim \rm DM_{IGM}/855 pc \ cm^{-3}
\end{equation}
at the redshift of the host galaxy of FRB 20240114A $z=0.1306$, and the derived \dmcosmic\ is about 111 pc cm$^{-3}$. 
All of these leave an upper limit of \dmhost of 321.3 pc cm$^{-3}$ (NE2001) or 332.2 pc cm$^{-3}$ (YMW16) at the observer frame, which includes contributions from both the host galaxy and the local environment of the FRB.
 
We use the extinction-corrected \citep{Cardelli1989ApJ...345..245C} H$\alpha$ emission line and its emission measure to estimate the DM contribution from the ionized gas of the host galaxy \citep{Cordes2016arXiv160505890C, Tendulkar121102-2017ApJ...834L...7T, OckerHost2022ApJ...931...87O}. 
To estimate the averaged \Ha\ emission surface flux density, we adopt a $a=0.5^{\prime \prime}$ and $b=0.4^{\prime \prime}$ ellipse since this galaxy is not resolved in Pan-STARRS \textit{i}-band image which covers the redshifted \Ha\ emission.
The \Ha\ surface density is $S_{\rm H\alpha} \approx (6.6 \pm 0.7) \times 10^{-16}$ erg cm$^{-2}$ s$^{-1}$ arcsec$^{-2}$ (or $ 120 \pm 10 $ Rayleigh ) in the rest frame, implying the emission measure is \citep{1977ApJ...216..433R}

\begin{equation}
    \begin{aligned}
        \rm{EM_{\mathrm{H\alpha}}^{s}} &= 2.75 \, \mathrm{pc \ cm}^{-6} \, T_{4}^{0.9} \, S({\mathrm{H}\alpha})
        \\
        &\approx 318 \pm 33 \ {\rm pc \ cm^{-6}} \times T_{4}^{0.9}, 
    \end{aligned}
\end{equation}
where $T_4$ is the temperature in units of $10^4$ K.
The \Ha\ contribution from ionized gas in \Ha\ emission regions to the DM budget can be expressed as
\begin{equation}
    \begin{aligned}
            \rm{DM_{H\alpha}^s} &= \rm{EM_{H\alpha}^{1/2}} \cdot \it{l}^{\rm 1/2} \cdot \left [ \frac{\zeta(\rm 1+\epsilon^2)}{f} \right ]^{\rm -1/2}
            \\
            &\approx 18 \pm 1 \ {\rm pc \ cm^{-3}} T_{4}^{0.45}  l^{1/2} \left [ \frac{\zeta(1+\epsilon^2)}{f} \right ]^{-1/2},
    \end{aligned}
\end{equation}
where $l$ represents the path length through the ionized gas. 
The parameters $\zeta$, $\epsilon ^2$ and $f$ are the ionized cloudlet model parameters that represent the cloud-cloud variations in the mean density ($\zeta \geq 1$), the variance of density fluctuations in a cloud ($0 \leq \epsilon ^2 \leq 1$), and the filling factor ($0 \leq f \leq 1$), respectively.

We use the elongated scale of the host galaxy, defined as $\rm 2a = 1^{\prime \prime}$, corresponding to $l = 2.4 \, \rm kpc$, to estimate the DM contribution from the ionized interstellar medium (ISM).  
90\% or more of the ionized ISM resides in the diffuse ionized gas (DIG, or warm ionized medium, WIM, in the case of the Milky Way), which surrounds the galaxy and is primarily traced by \Ha\ emission \citep{WIM2009RvMP...81..969H}.  
For the estimation, we adopt the typical values of the Milky Way's WIM, namely $f = 0.1$, $\zeta = 1$, and $\epsilon^2 = 1$.
We use $\rm T_4=1$ for the electron temperature of an H\textsc{ii} region \citep{DraineISM2011piim.book.....D}.
The estimated \Ha-traced DM$_{\rm host}$ on the source frame is  
DM$_{\rm{H\alpha}}^{s} = 195 \pm 10 \ {\rm pc \ cm^{-3}}$. 
Divieded by the factor $(1+z)$, the observed DM$_{\rm host}$ is
DM$_{\rm host} = 173 \pm 9 \ {\rm pc \ cm^{-3}}$.
In the range of $1 \leq \zeta(1+\epsilon^2)/f \leq 50$ \citep{OckerHost2022ApJ...931...87O}, the observer frame DM$_{\rm host}$ could be as high as 
$ 773 \pm 40 \ {\rm pc \ cm^{-3}}$ or as low as $ 109 \pm 6 \ {\rm pc \ cm^{-3}}$.

The estimation shows the ionized ISM of the host galaxy of FRB 20240114A can contribute considerably to the \dmhost.
The distribution of ionized gas in the star formation regions in the host galaxy in the line of sight may increase DM contribution estimate.
In the star-forming dwarf galaxies, due to the photoionization by massive stars and the porous nature of the ISM, diffuse ionized gas (DIG) is present outside the boundaries of the H\textsc{ii} regions, akin to spiral galaxies \citep{Martin1997, Hidalgo-G´amez2007, Polles2019}.
The brightest H\textsc{ii} regions in irregular dwarf galaxies span sizes ranging from tens to hundreds of parsecs, while the distribution of DIG extends to approximately 1 kiloparsec \citep{Martin1997}.

\subsection{SED Fitting and Stellar Mass}
\label{subsect:sed}

To decompose the stellar components in the host galaxy, we follow the methodology described by \citet{2015ApJ...804...27A}, as also employed by \citet{2023ApJ...958..162L}, to model the UV-optical SED. 
Specifically, the SED of a galaxy is a linear combination of four empirical spectral templates \citep{2010ApJ...713..970A}, three of which correspond to galaxy SED templates (``E” galaxy for an old stellar population, an intermediate star-forming ``Sbc” galaxy, and a starburst ``Im” galaxy) and one corresponding to an unreddened AGN.
Although the absence of broad emission line and the BPT analysis (see Sect.~\ref{subsect:spectrum}) indicate no AGN activity in this host dwarf galaxy, we consider an obscured AGN component in the SED fitting despite of its rareness in the low-mass galaxy system. 
We adopted $R_{\rm V}$=3.1 and an SMC reddening law \citep{2003ApJ...594..279G} at short wavelengths and a Milky Way reddening law \citep{2007ApJ...663..320F} at longer wavelengths (see \citealp{2010ApJ...713..970A} for further details). 

Following \cite{2010ApJ...713..970A}, we fit the SED using the Markov Chain Monte Carlo (MCMC) implementation of \citet{emcmc2013PASP..125..306F} through the public python package emcee, solving for the amplitude of the three template components and the AGN SED template with obscuration. 
We use uninformative priors for all parameters, enforcing all to be non-negative. 
The median of the marginalized distribution of each parameter is taken as its best-fit value, and the quoted 1$-\sigma$ uncertainties correspond to the 16th and 84th percentiles of the distributions. 
The flux of each band was corrected for extinction from dust in the Milky Way before modeling process \citep[$A_{\rm V}$=0.18 mag;][]{1998ApJ...500..525S,2007ApJ...663..320F,2011ApJ...737..103S}.

\begin{figure}[htbp]
\begin{center}
\epsscale{1.1}
\plotone{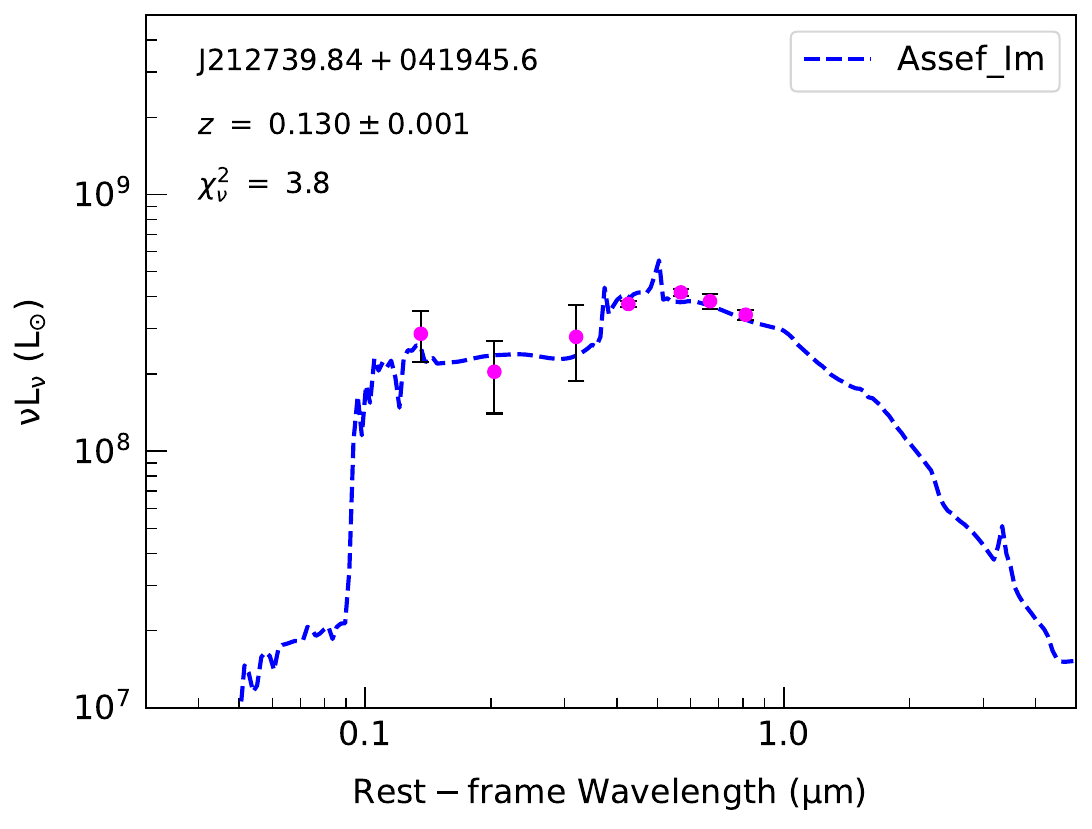}
\end{center}
\caption{Best-fit SED template model of the host galaxy of FRB 20240114A. 
The magenta points with error bars represent the extinction-corrected photometric data. 
The SED modeling uses the template of the irregular galaxy from \citet{2010ApJ...713..970A} as described in Section~\ref{subsect:sed}.
\label{fig:fig1}}
\end{figure}

Figure \ref{fig:fig1} shows the best-fit SED model of the host galaxy of FRB 20240114A. 
The best fit model indicate that the ``Im” template dominate its UV-optical SED and other components (“E”, “Sbc”, and “AGN”) were negligible. 
The F-test, which assesses the improvement in $\chi^{2}$ from adding other templates and shows a $\leq$10\% chance of being spurious, also yielded a similar result. 
This suggests that the host galaxy is likely a star-forming system, as indicated by the BPT diagram (Figure~\ref{fig:bpt}).
We estimate its optical luminosity blueward of 1$\micron$ ($L_{\rm optical}$) to be $6.8\times 10^{8} L_{\odot}$ by integrating over the SED (Figure \ref{fig:fig1}), using power-law interpolation on the best-fit SED model \citep{2023ApJ...958..162L}.
The star formation rate derived using the fitted UV continuum is SFR(UV)$\sim 0.09 M_{\odot} \ \rm yr^{-1}$. 
A detailed discussion of SFR estimation and its application is provided in Sect.~\ref{sect:mass_and_SFMS}.

Table~\ref{tab:gal} summarizes all the derived properties of the host galaxy.
The stellar mass ($M_{\rm star}$) is often derived from near-infrared (NIR) photometry, as it traces the low-mass stars that dominate host galaxies and is less affected by extinction. 
Due to the lack of NIR data for the host galaxy of FRB 20240114A, we use the correlations between \textit{ugriz} colors and M/L ratios reported by \citet{2003ApJS..149..289B} to estimate its stellar mass. 
The M/L ratios are given by ${{\log}_{10}(M/L_{\rm \nu})=\rm{a_{\nu}}+(\rm{b_{\nu}}\times color)}$, where $L_{\rm \nu}$ is the monochromatic luminosity in optical bands (\textit{g}, \textit{r}, and \textit{i}), and the calibrated parameters ($\rm a_{\nu}$ and $\rm b_{\nu}$) are provided in Table~7 of \citet{2003ApJS..149..289B}. 
Specifically, adopting a stellar initial mass function (IMF) from \citet{1955ApJ...121..161S}, we use the best-fit SED model shown in Figure \ref{fig:fig1} to estimate the rest-frame optical colors and luminosities.
The estimated $M_{\rm star}$ is $10^{8.6 \pm 0.2} \ M_{\odot}$ for the host galaxy of FRB 20240114A, with the uncertainty based on the errors of the magnitudes. 

\begin{table}[htbp]
    \centering
    \caption{Properties of the host galaxy of FRB 20240114A.
    }
    \medskip
    \begin{threeparttable}
    \begin{tabular*}{0.85\linewidth}{lc}
    \hline
    \hline
    \noalign{\smallskip}
         Redshift & 0.1306 $\pm$ 0.0002 \\
         \noalign{\smallskip}
         Stellar Mass ($\mathcal{M}_{\odot}$) & $(4.0 \pm 1.8) \times 10^8 $ \\
         \noalign{\smallskip}
         $12 + \rm log_{10}([O/H])$ (dex) & $ \sim 8.5 $ \\
         \noalign{\smallskip}
         $L_{\rm{H\alpha}}$ (erg s$^{-1}$) & $(1.2 \pm 0.1) \times 10^{40} $ \\
         \noalign{\smallskip}
         SFR(H${\alpha}$) ($\mathcal{M}_{\odot} \ {\rm yr}^{-1}$)  & $ 0.06 \pm 0.01 $ \\
         \noalign{\smallskip}
         log$_{10}$ sSFR(H${\alpha}$) (yr$^{-1}$) & $-9.8 \pm 0.2$ \\
         \noalign{\smallskip}
         $ S({\rm H \alpha})_s^{\ast}$ & $(6.6 \pm 0.7) \times 10^{-16}$ \\
         \noalign{\smallskip}
         DM(H$\alpha$)$_s$ (pc cm $^{-3}$)$^{\diamond} $ & $200 \pm 10$ \\
         \noalign{\smallskip}
         \hline
    \end{tabular*}
    \medskip
    \begin{tablenotes}
    \footnotesize 
    \item[$\ast$] Assuming the galaxy has a elliptical size with $a=0.5 ^{\prime \prime}$, $b=0.4^{\prime \prime}$. The unit of $ S({\rm H \alpha})_s $ is erg cm$^{-2}$ s$^{-1}$ arcsec$^{-2}$. $ S({\rm H \alpha})_s $ is equal to $120 \pm 10$ Rayleigh.
    \item[$\diamond$] Assuming $L=2.4$ kpc, $f=0.1$, $\zeta=1$ and $\epsilon^2=1$.
    \end{tablenotes}
    \end{threeparttable}
    \label{tab:gal}
\end{table}

\section{Discussion}
\label{sect:dissc}

\subsection{Host Glaxsy's SFR and Comparison to SF Main Sequence} 
\label{sect:mass_and_SFMS}

The UV continuum of galaxies traces young, massive stars and is widely used as an indicator of SFR \citep[e.g.,][]{1998ARA&A..36..189K}. 
For the host galaxy, the best-fit model indicates that the UV-optical SED is dominated by a starburst component, making it a good tracer of the SFR.
However, dust attenuation, especially at UV wavelength, can weaken its effectiveness in measuring the SFR. 
In nearby star-forming galaxies, UV dust attenuation can vary over several magnitudes \citep[e.g.,][]{2005ApJ...619L..51B,2005MNRAS.360.1413B}. 
We use the fitted infrared luminosity of FRB 20240114A's host galaxy to correct for the dust attenuation in the UV
and calculate the dust-corrected SFR \citep[e.g.,][]{2012ARA&A..50..531K}. 
By integrating over the best-fit SED in the $1500-2800$ \AA\ range, the average UV luminosity ($L_{\nu}$) of $6.5\times 10^{26} \ \rm{erg \ s^{-1} \ Hz^{-1}}$ implies an SFR of $\sim$ 0.09 $M_{\odot} \ \rm{yr^{-1}}$ using a Salpeter initial mass function (IMF) for a stellar population with stars ranging from 0.1 to 100 $M_{\odot}$ \citep[][Equation~(1)]{1998ARA&A..36..189K}. 
However, as dust correction relies on infrared luminosity, the SFR derived from the UV continuum without explicit dust attenuation correction should be considered a lower limit.
Given that \Ha\ emission preferentially traces younger stellar populations compared to UV emission \citep{2012ARA&A..50..531K}, we utilize SFR(H${\alpha}$) in subsequent analyses.

\begin{figure*}[htb]
    \centering
    \includegraphics[width=0.7\linewidth]{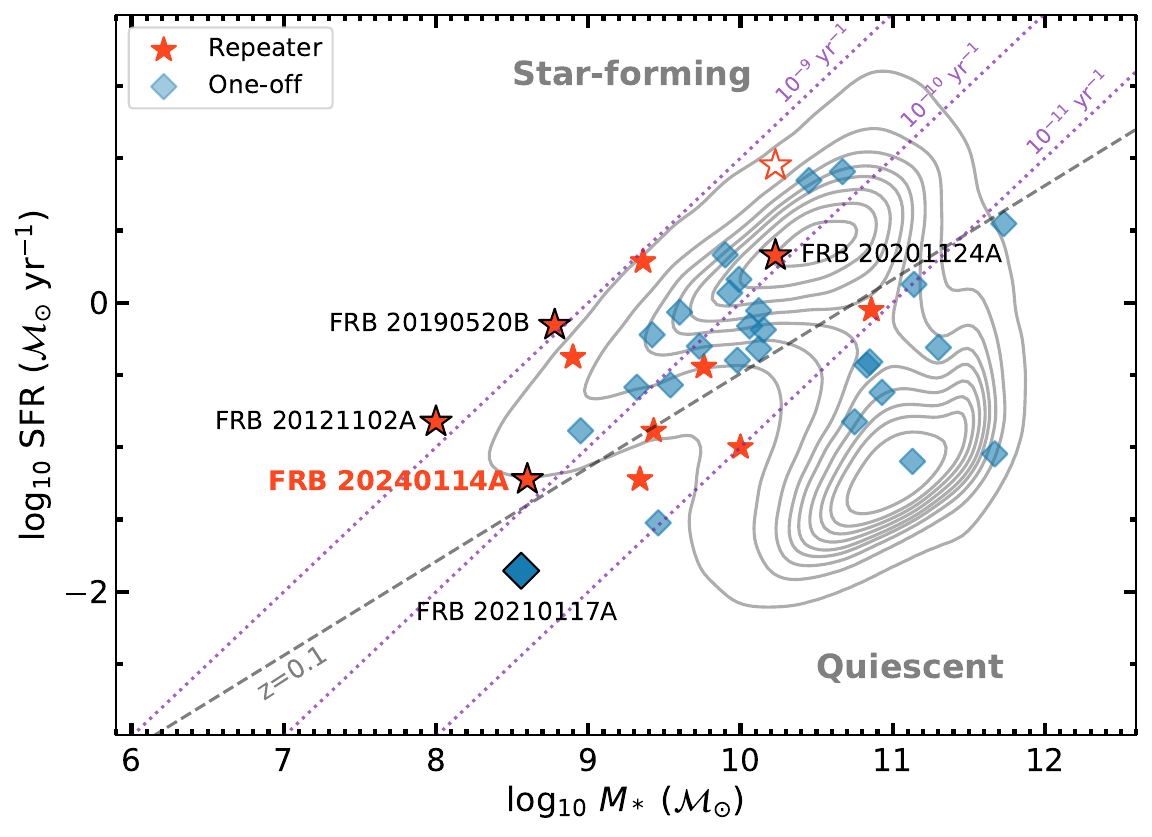}
    \caption{SFR and stellar mass of FRB host galaxies.
    The host galaxy of FRB 20240114A, other dwarf host galaxies, and the host galaxy of the active repeater FRB 20201124A are highlighted with larger symbols outlined in black, with their names labeled near the symbols.
    The unfilled red star represents the SFR(radio) of the host galaxy of FRB 20201124A \citep{Dong201124A2024ApJ...961...44D}.
    The dashed gray line indicates the star-forming main sequence at redshift $z=0.1$.
    The dotted purple lines denote equal-sSFR levels of sSFR$= 10^{-9}, 10^{-10} \rm \ and \ 10^{-11}$.
    The contours represent SDSS galaxies \citep[DR7;][]{2009ApJS..182..543A} with redshifts between 0.01 and 0.2, provided by the MPA-JHU catalog.}
    \label{fig:SFR}
\end{figure*}

Figure~\ref{fig:SFR} shows the star formation rate and the stellar mass of the host galaxies of repeating and one-off FRBs, along with properties of galaxies with redshifts between 0.01 and 0.2 in the Max Planck Institute for Astrophysics-Johns Hopkins University (MPA-JHU) value-added catalog\footnote{\url{https://wwwmpa.mpa-garching.mpg.de/SDSS/DR7/}}, which is based on the SDSS DR7 spectra \citep{2009ApJS..182..543A}. 
The SFR(\Ha) value of FRB 20240114's host galaxy is used to facilitate comparison with the SFR(\Ha) of the host galaxies of FRB 20121102A and FRB 20190520B.
The host galaxy of FRB 20240114A lies on the star forming main sequence (SFMS).
However, compared to the dwarf host galaxies of the repeating FRBs FRB 20121102A and FRB 20190520B, its SFR is lower, making it more similar to the dwarf host galaxy of the one-off FRB 20210117A. 
It is noteworthy that FRB 20210117A exhibits a distinctive downward-drifting structure in its dynamic spectrum, a characteristic now recognized as a hallmark of repeating FRBs.

From Figure~\ref{fig:SFR}, we found that the properties of the host galaxies of repeating and one-off FRBs exhibits some differences in their distribution. 
To quantitatively assess the significance of these differences, we use the MANOVA test \citep{MANOVAanderson2009introduction}, available through the Python package\footnote{\url{https://www.statsmodels.org/stable/generated/statsmodels.multivariate.manova.MANOVA.html}}.
MANOVA assumes the data from all groups have common variance, so that the variability in the data does not depend on group membership.
Our data includes 11 host galaxies of repeaters with redshifts up to 0.6 and 33 host galaxies of one-off FRBs with redshifts up to 1.
The subjects are independently sampled and the data are normally distributed.
The MANOVA test yields a $p$-value of 0.0116, which is below the common significance threshold of $p=0.05$, indicating a significant difference in the stellar mass and SFR distributions between the host galaxies of repeating and one-off FRBs.
Since the SFMS evolves with redshift \cite{SFMSredshift2023MNRAS.519.1526P}, the test results could be influenced by the differences in the redshift distributions between the two datasets.
Therefore, we removed the samples with redshifts higher than 0.6 from the host galaxies of one-off FRBs, and the new MANOVA $p$-value is 0.0203, still below the threshold of 0.05.
Our results suggest potential differences in the stellar populations of the progenitors of repeating versus one-off FRBs. 
Previous studies by \cite{Gordon2023hostgalaxies} and \cite{Sharma2024arXiv240916964S} found no statistical differences between the host galaxies of repeaters and one-off FRBs, using the Anderson-Darling (AD) test and the Kolmogorov-Smirnov (KS) test, respectively, to assess individual properties such as stellar mass or star formation rate.
Our results, however, suggest a more intricate relationship between stellar mass and SFR that may not be evident when these properties are analyzed independently. 
Further details are discussed in our subsequent work (Chen et al., submitted to ApJ).

\subsection{Dwarf Host Galaxies and Progenitors}
\label{subsect:dwarf_hosts}

Other very active FRBs, including FRB 20121102A \citep{Tendulkar121102-2017ApJ...834L...7T, Bassa_2017_121102HST, Kokubo_2017_121102IFU}, FRB 20190520B \citep[][Chen et al. submitted to ApJ]{Niu190520-2022Natur.606..873N}, and FRB 20201124A \citep{201124AFong_2021, 201124APiro2021, Dong201124A2024ApJ...961...44D} are all associated with star-forming regions in their host galaxies. 
High spacial resolution observations show that FRB 20121102A is associated with the star-forming region and the \Ha\ emission region. 
The emission region in the host galaxy of FRB 20190520B is off-center and corresponds to the location of the FRB. 
The \Ha -traced star formation in the host galaxy of FRB 20190520B is concentrated in the emission region, which is suggested to be dominated by star formation according to the BPT diagram.
Although the overall galaxy sSFR for the host galaxy of FRB 20201124A, calculated using \Ha\ emission, is lower than that of the other two FRBs (Figure~\ref{fig:SFR}), infrared and radio observations reveal a highly obscured star-forming region associated with the FRB. 
These three FRBs, namely FRB 20121102A, FRB 20190520B and FRB 20201124A, are suggested to have young stellar progenitors associated with star forming \citep{Bassa_2017_121102HST, FengRM2022Sci...375.1266F, Dong201124A2024ApJ...961...44D}.
Further observations of the local environment of FRB 20240114A, as well as the star-forming regions in its host galaxy, will provide additional constraints on the formation mechanisms of this FRB.

\subsection{Foreground Cluster}
\label{subsect:group_and_clusters}

\begin{figure}[!ht]
    \centering
    \includegraphics[width=\linewidth]{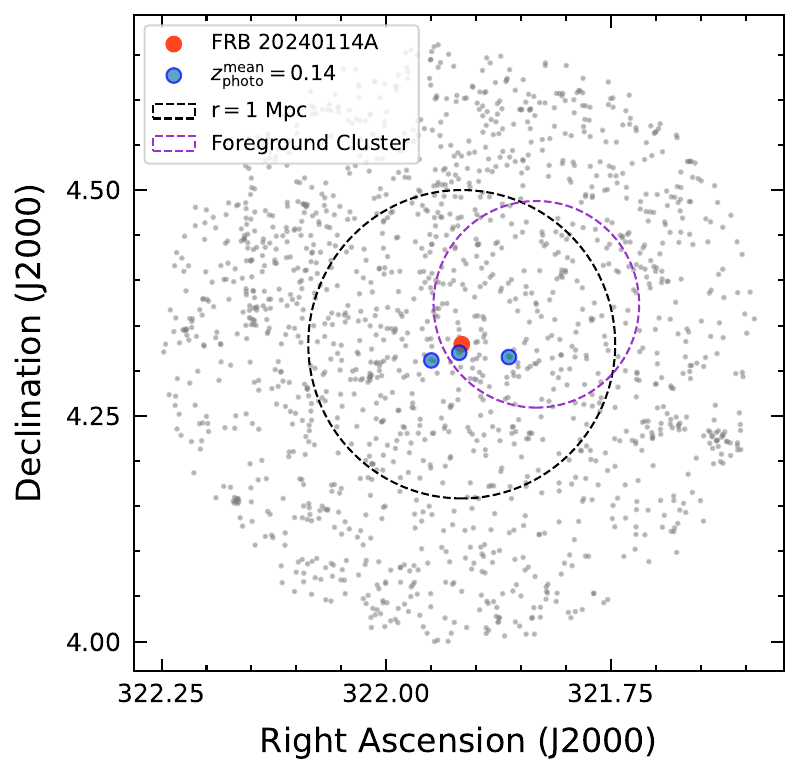}
    \caption{Locations of FRB 20240114A (red dot) and the identified galaxy group candidate (blue dots).
    The black dashed circle represents the typical scale of a galaxy cluster with a radius of 1 Mpc.
    The remaining gray dots represent other galaxies used in the FoF algorithm analysis that are not part of the identified group.
    The projection of $r_{500}$ for the foreground galaxy cluster is indicated by the purple dashed circle.
    }
    \label{fig:group}
\end{figure}

We analyze the nearby galaxies of the host galaxy of FRB 20240114A for the potential foreground galaxy group or cluster candidates \citep{2024ATel16613....1B} within $20^{\prime}$ from FRB 20240114A.
We use the spectroscopic redshift of the host galaxy to analyze potential clustering.
However, since the photometric redshift (photo-\emph{z}) of the host galaxy covers a broader range \citep[$0.252 \pm 0.124$, ][]{LS2019AJ....157..168D}) due to its color and faintness, we extend the redshift range to $0 \sim 0.3$ to account for possible color-related effects in the dataset used for the search.
We applied the fiend-of-friend (FoF) group-finding algorithm \citep{Turner1976, Press1982} around FRB 20240114A, utilizing photo-\emph{z} data from the Legacy Survey DR9.
The basic steps of this method involve sorting galaxies within a specified linking length ($LL$) into groups or clusters. 
According to the fitted scaling relation of SDSS samples, $LL$ follows an arctan law as a function of redshift \citep{tempel2012, tempel2014}:
 \begin{equation}
    d_{LL}(z) = d_{LL,0} \left[ 1 + a \arctan\left(\frac{z}{z_*}\right) \right]
    \label{LL function}
\end{equation}
where $d_{LL,0}$ is $LL$ at $ z = 0 $, and a and $z_*$ are free parameters. 
We set $d_{LL,0}$ to 0.35 h$^{-1}$ Mpc and applied this method within a radius of 20$^{\prime}$. 
By controlling the group number density in the range of 0.001-0.005 $(h^{-1} \, \text{Mpc})^{-3}$ at $z \lesssim 0.13$ \citep{tempel2014}, we determined the final parameters as $a = 5.0$ and ${z_*} = 0.05$ to identify groups with a richness (i.e., the number of group members) greater than 2, as shown in Figure~\ref{fig:group}.
This group could be potentially associated with the FRB host galaxy given its photo-\emph{z}.
The high surface density of galaxies surrounding FRB 20240114A suggests a relatively dense galaxy environment.

Furthermore, we checked the catalog of galaxy clusters from the sky surveys within a 1 Mpc radius centered around FRB 20240114A. 
We found no matched X-ray cluster in the catalogs of BAX database \citep{Sadat2004A&A...424.1097S}, MCXC database \citep{MCXC2011A&A...534A.109P}, NORAS \& REFLEX \citep{NORAS2000ApJS..129..435B, REFLEX2004A&A...425..367B}, and eROSITA \citep{eRosita2024A&A...685A.106B}.
However, one foreground cluster, J212719.9$+$042225, was identified using the catalog from \cite{Wen2024ApJS..272...39W}, based on data from the DESI Legacy Imaging Survey.
The angular distance from the host galaxy of FRB 20240114A to the center of the foreground cluster J212719.9$+$042225 is $\theta=5.6^{\prime}$. 
The spectroscopic redshift of this galaxy cluster is $z=0.0903$, with a derived total mass of M$_{\rm 500} = 6.8 \times 10^{13} \ M_{\odot}$ and a radius of $r_{\rm 500}=0.67 \rm \ Mpc$.
The FRB is located within the $r_{500}$ radius of this galaxy cluste, suggesting a significant likelihood that the cluster contributes to the observed DM of FRB 20240114A. 
We estimate the electron density at $r_{\rm 500}$ to be $n_e \sim 0.0003$ cm$^{-3}$ \citep{clusterhalo2019ApJ...875...26F}.
Using a line-of-sight path length through the cluster of $l \sim 0.58$ Mpc, the contribution of this foreground cluster, $\rm DM_{fore}$, is $\sim 180$ pc cm$^{-3}$, which could make a non-negligible contribution to the total observed DM.
A similar situation was proposed by \cite{190520foreground2023ApJ...954L...7L} for the extremely high extra-galactic DM of FRB 20190520B.

In Sect.~\ref{subsect:halphaDM}, we obtained a DM surplus of approximately 330 pc cm$^{-3}$ after subtracting the contributions from the Milky Way and the IGM. 
Subtracting the contribution from the foreground galaxy cluster, the remaining \dmhost\ is about 150 pc cm$^{-3}$, which is roughly 50 pc cm$^{-3}$ lower than the estimate of the host galaxy's contribution obtained using the \Ha\ emission line (Table~\ref{tab:gal}).
The discrepancy could be caused by the inhomogeneous distribution or clumpiness of the ionized gas in the host galaxy, the use of simplified Milky Way WIM parameters, or an incomplete analysis of the foreground contributions.  
Further observations will be required to resolve this inconsistency.

\section{Summary}
\label{sect:summary}

In this study, we present the optical spectroscopic observations of the host galaxy of the hyper-active FRB 20240114A. 
Emission lines of \Ha, \Hb, [\OIII]$\lambda\lambda$4959,5007, [\NII]$\lambda\lambda$6548,6583, and [\SII]$\lambda$6716 are detected. 
Both the BPT diagram and the SED fitting suggest that these emission lines are dominated by star forming in the host galaxy.
Combining our spectroscopic data with archived data from large surveys, we determine that this host galaxy is a low-metallicity dwarf galaxy 
($12+ {\rm log_{10}([O/H])} \sim 8.5 $) at a redshift of $z=0.1306 \pm 0.0002$. 
The intrinsic extinction of the host galaxy is $A_v = 0.22$. 
The stellar mass derived from the \textit{ugriz} colors is $(4.0 \pm 1.8 )\times 10^8 \ M_{\odot}$.
The star formation rate derived from the extinction corrected \Ha\ luminosity is SFR(\Ha)$\sim 0.06 \ M_{\odot} \rm{yr^{-1}}$.%
We use the emission measure of \Ha\ to estimate the dispersion measure contribution of the ionized ISM in the host galaxy. 
With DM$_{\rm{host}} = (200 \pm 10) \times T_4$ pc cm $^{-3}$, the ionized ISM of the host galaxy can potentially contribute significantly to the extra-galactic DM of FRB 20240114A.
We confirmed the presence of the foreground galaxy cluster by consulting the published catalogs and estimated its DM contribution to be $\sim 180$ pc cm$^{-3}$.
Additionally, our analysis of nearby galaxies indicates that the host galaxy of FRB 20240114A is situated in a relatively dense environment.

The similarity among the star-forming dwarf host galaxies of FRB 20121102A, FRB 20190520B, and FRB 20240114A, along with their high activity levels, suggests that these FRBs may share a common population or formation mechanism.
The high correspondence between the star-forming regions in the host galaxies and the locations of the actively repeating FRBs, such as FRB 20121102A, FRB 20190520B, and FRB 20201124A, may provide insights into the progenitor population of these FRBs. 
Follow-up observations of the local environment of FRB 20240114A, along with further analysis of the correlated PRS candidate observed by \cite{2024ATel16695....1Z} and \cite{uGMRT2024arXiv241213121B}, are crucial for revealing the formation environment and evolutionary state of this hyper-active FRB.

We also analyzed the distribution of FRB host galaxies on the stellar mass versus SFR diagram. 
The results of the MANOVA test indicate that the distributions of the host galaxies of repeating and one-off FRBs are significantly different, with a $p$-value of 0.0116.
This suggests that the formation environments and mechanisms of repeaters and one-offs could be statistically distinct based on these distributions.

C.-W.T., X.L.C. and P.W. acknowledge support from the National Natural Science Foundation of China (NSFC) Programs No. 11988101, No. 12041302.
The authors would like to thank the reviewer for their useful comments on the manuscript.
This work is supported by CAS project No. JZHKYPT-2021-06.
Di Li is a New Cornerstone Investigator.
P.W. acknowledges support from the CAS Youth Interdisciplinary Team, the Youth Innovation Promotion Asbasociation CAS (id. 2021055), and the Cultivation Project for FAST Scientific Payoff and Research Achievement of CAMS-CAS.
M.L. is supported by the International Partnership Program of Chinese Academy of Sciences, Program No.114A11KYSB20210010.
We thank Antonio Luis Cabrera Lavers for his help on observations using GTC.

This work has used the data from the Five-hundred-meter Aperture Spherical radio Telescope (FAST).  FAST is a Chinese national mega-science facility, operated by the National Astronomical Observatories of Chinese Academy of Sciences (NAOC).

Based on observations made with the Gran Telescopio Canarias (GTC), installed at the Spanish Observatorio del Roque de los Muchachos of the Instituto de Astrofísica de Canarias, on the island of La Palma.

This work is (partly) based on data obtained with the instrument OSIRIS, built by a Consortium led by the Instituto de Astrofísica de Canarias in collaboration with the Instituto de Astronomía of the Universidad Autónoma de México. OSIRIS was funded by GRANTECAN and the National Plan of Astronomy and Astrophysics of the Spanish Government.

\facilities{FAST, GTC (OSIRIS+)}

\software{
Astropy \citep{Astropy2022ApJ...935..167A},
emcmc \citep{emcmc2013PASP..125..306F},
PyPeit \citep{pypeit1, pypeit2}, 
SciPy \citep{2020SciPy-NMeth}
}

\bibliography{sample631}{}

\begin{thebibliography}{}
\expandafter\ifx\csname natexlab\endcsname\relax\def\natexlab#1{#1}\fi
\providecommand{\url}[1]{\href{#1}{#1}}
\providecommand{\dodoi}[1]{doi:~\href{http://doi.org/#1}{\nolinkurl{#1}}}
\providecommand{\doeprint}[1]{\href{http://ascl.net/#1}{\nolinkurl{http://ascl.net/#1}}}
\providecommand{\doarXiv}[1]{\href{https://arxiv.org/abs/#1}{\nolinkurl{https://arxiv.org/abs/#1}}}

\bibitem[{{Abazajian} {et~al.}(2009){Abazajian}, {Adelman-McCarthy}, {Ag{\"u}eros}, {Allam}, {Allende Prieto}, {An}, {Anderson}, {Anderson}, {Annis}, {Bahcall}, {Bailer-Jones}, {Barentine}, {Bassett}, {Becker}, {Beers}, {Bell}, {Belokurov}, {Berlind}, {Berman}, {Bernardi}, {Bickerton}, {Bizyaev}, {Blakeslee}, {Blanton}, {Bochanski}, {Boroski}, {Brewington}, {Brinchmann}, {Brinkmann}, {Brunner}, {Budav{\'a}ri}, {Carey}, {Carliles}, {Carr}, {Castander}, {Cinabro}, {Connolly}, {Csabai}, {Cunha}, {Czarapata}, {Davenport}, {de Haas}, {Dilday}, {Doi}, {Eisenstein}, {Evans}, {Evans}, {Fan}, {Friedman}, {Frieman}, {Fukugita}, {G{\"a}nsicke}, {Gates}, {Gillespie}, {Gilmore}, {Gonzalez}, {Gonzalez}, {Grebel}, {Gunn}, {Gy{\"o}ry}, {Hall}, {Harding}, {Harris}, {Harvanek}, {Hawley}, {Hayes}, {Heckman}, {Hendry}, {Hennessy}, {Hindsley}, {Hoblitt}, {Hogan}, {Hogg}, {Holtzman}, {Hyde}, {Ichikawa}, {Ichikawa}, {Im}, {Ivezi{\'c}}, {Jester}, {Jiang}, {Johnson}, {Jorgensen}, {Juri{\'c}}, {Kent}, {Kessler}, {Kleinman}, {Knapp},
  {Konishi}, {Kron}, {Krzesinski}, {Kuropatkin}, {Lampeitl}, {Lebedeva}, {Lee}, {Lee}, {French Leger}, {L{\'e}pine}, {Li}, {Lima}, {Lin}, {Long}, {Loomis}, {Loveday}, {Lupton}, {Magnier}, {Malanushenko}, {Malanushenko}, {Mandelbaum}, {Margon}, {Marriner}, {Mart{\'\i}nez-Delgado}, {Matsubara}, {McGehee}, {McKay}, {Meiksin}, {Morrison}, {Mullally}, {Munn}, {Murphy}, {Nash}, {Nebot}, {Neilsen}, {Newberg}, {Newman}, {Nichol}, {Nicinski}, {Nieto-Santisteban}, {Nitta}, {Okamura}, {Oravetz}, {Ostriker}, {Owen}, {Padmanabhan}, {Pan}, {Park}, {Pauls}, {Peoples}, {Percival}, {Pier}, {Pope}, {Pourbaix}, {Price}, {Purger}, {Quinn}, {Raddick}, {Re Fiorentin}, {Richards}, {Richmond}, {Riess}, {Rix}, {Rockosi}, {Sako}, {Schlegel}, {Schneider}, {Scholz}, {Schreiber}, {Schwope}, {Seljak}, {Sesar}, {Sheldon}, {Shimasaku}, {Sibley}, {Simmons}, {Sivarani}, {Allyn Smith}, {Smith}, {Smol{\v{c}}i{\'c}}, {Snedden}, {Stebbins}, {Steinmetz}, {Stoughton}, {Strauss}, {SubbaRao}, {Suto}, {Szalay}, {Szapudi}, {Szkody}, {Tanaka},
  {Tegmark}, {Teodoro}, {Thakar}, {Tremonti}, {Tucker}, {Uomoto}, {Vanden Berk}, {Vandenberg}, {Vidrih}, {Vogeley}, {Voges}, {Vogt}, {Wadadekar}, {Watters}, {Weinberg}, {West}, {White}, {Wilhite}, {Wonders}, {Yanny}, {Yocum}, {York}, {Zehavi}, {Zibetti}, \& {Zucker}}]{2009ApJS..182..543A}
{Abazajian}, K.~N., {Adelman-McCarthy}, J.~K., {Ag{\"u}eros}, M.~A., {et~al.} 2009, \apjs, 182, 543, \dodoi{10.1088/0067-0049/182/2/543}

\bibitem[{{Abdurro'uf} {et~al.}(2022){Abdurro'uf}, {Accetta}, {Aerts}, {Silva Aguirre}, {Ahumada}, {Ajgaonkar}, {Filiz Ak}, {Alam}, {Allende Prieto}, {Almeida}, {Anders}, {Anderson}, {Andrews}, {Anguiano}, {Aquino-Ort{\'\i}z}, {Arag{\'o}n-Salamanca}, {Argudo-Fern{\'a}ndez}, {Ata}, {Aubert}, {Avila-Reese}, {Badenes}, {Barb{\'a}}, {Barger}, {Barrera-Ballesteros}, {Beaton}, {Beers}, {Belfiore}, {Bender}, {Bernardi}, {Bershady}, {Beutler}, {Bidin}, {Bird}, {Bizyaev}, {Blanc}, {Blanton}, {Boardman}, {Bolton}, {Boquien}, {Borissova}, {Bovy}, {Brandt}, {Brown}, {Brownstein}, {Brusa}, {Buchner}, {Bundy}, {Burchett}, {Bureau}, {Burgasser}, {Cabang}, {Campbell}, {Cappellari}, {Carlberg}, {Wanderley}, {Carrera}, {Cash}, {Chen}, {Chen}, {Cherinka}, {Chiappini}, {Choi}, {Chojnowski}, {Chung}, {Clerc}, {Cohen}, {Comerford}, {Comparat}, {da Costa}, {Covey}, {Crane}, {Cruz-Gonzalez}, {Culhane}, {Cunha}, {Dai}, {Damke}, {Darling}, {Davidson}, {Davies}, {Dawson}, {De Lee}, {Diamond-Stanic}, {Cano-D{\'\i}az}, {S{\'a}nchez},
  {Donor}, {Duckworth}, {Dwelly}, {Eisenstein}, {Elsworth}, {Emsellem}, {Eracleous}, {Escoffier}, {Fan}, {Farr}, {Feng}, {Fern{\'a}ndez-Trincado}, {Feuillet}, {Filipp}, {Fillingham}, {Frinchaboy}, {Fromenteau}, {Galbany}, {Garc{\'\i}a}, {Garc{\'\i}a-Hern{\'a}ndez}, {Ge}, {Geisler}, {Gelfand}, {G{\'e}ron}, {Gibson}, {Goddy}, {Godoy-Rivera}, {Grabowski}, {Green}, {Greener}, {Grier}, {Griffith}, {Guo}, {Guy}, {Hadjara}, {Harding}, {Hasselquist}, {Hayes}, {Hearty}, {Hern{\'a}ndez}, {Hill}, {Hogg}, {Holtzman}, {Horta}, {Hsieh}, {Hsu}, {Hsu}, {Huber}, {Huertas-Company}, {Hutchinson}, {Hwang}, {Ibarra-Medel}, {Chitham}, {Ilha}, {Imig}, {Jaekle}, {Jayasinghe}, {Ji}, {Johnson}, {Jones}, {J{\"o}nsson}, {Katkov}, {Khalatyan}, {Kinemuchi}, {Kisku}, {Knapen}, {Kneib}, {Kollmeier}, {Kong}, {Kounkel}, {Kreckel}, {Krishnarao}, {Lacerna}, {Lane}, {Langgin}, {Lavender}, {Law}, {Lazarz}, {Leung}, {Leung}, {Lewis}, {Li}, {Li}, {Lian}, {Liang}, {Lin}, {Lin}, {Lin}, {Lintott}, {Long}, {Longa-Pe{\~n}a}, {L{\'o}pez-Cob{\'a}}, {Lu},
  {Lundgren}, {Luo}, {Mackereth}, {de la Macorra}, {Mahadevan}, {Majewski}, {Manchado}, {Mandeville}, {Maraston}, {Margalef-Bentabol}, {Masseron}, {Masters}, {Mathur}, {McDermid}, {Mckay}, {Merloni}, {Merrifield}, {Meszaros}, {Miglio}, {Di Mille}, {Minniti}, {Minsley}, {Monachesi}, {Moon}, {Mosser}, {Mulchaey}, {Muna}, {Mu{\~n}oz}, {Myers}, {Myers}, {Nadathur}, {Nair}, {Nandra}, {Neumann}, {Newman}, {Nidever}, {Nikakhtar}, {Nitschelm}, {O'Connell}, {Garma-Oehmichen}, {Luan Souza de Oliveira}, {Olney}, {Oravetz}, {Ortigoza-Urdaneta}, {Osorio}, {Otter}, {Pace}, {Padilla}, {Pan}, {Pan}, {Parikh}, {Parker}, {Peirani}, {Pe{\~n}a Ram{\'\i}rez}, {Penny}, {Percival}, {Perez-Fournon}, {Pinsonneault}, {Poidevin}, {Poovelil}, {Price-Whelan}, {B{\'a}rbara de Andrade Queiroz}, {Raddick}, {Ray}, {Rembold}, {Riddle}, {Riffel}, {Riffel}, {Rix}, {Robin}, {Rodr{\'\i}guez-Puebla}, {Roman-Lopes}, {Rom{\'a}n-Z{\'u}{\~n}iga}, {Rose}, {Ross}, {Rossi}, {Rubin}, {Salvato}, {S{\'a}nchez}, {S{\'a}nchez-Gallego}, {Sanderson}, {Santana
  Rojas}, {Sarceno}, {Sarmiento}, {Sayres}, {Sazonova}, {Schaefer}, {Schiavon}, {Schlegel}, {Schneider}, {Schultheis}, {Schwope}, {Serenelli}, {Serna}, {Shao}, {Shapiro}, {Sharma}, {Shen}, {Shetrone}, {Shu}, {Simon}, {Skrutskie}, {Smethurst}, {Smith}, {Sobeck}, {Spoo}, {Sprague}, {Stark}, {Stassun}, {Steinmetz}, {Stello}, {Stone-Martinez}, {Storchi-Bergmann}, {Stringfellow}, {Stutz}, {Su}, {Taghizadeh-Popp}, {Talbot}, {Tayar}, {Telles}, {Teske}, {Thakar}, {Theissen}, {Tkachenko}, {Thomas}, {Tojeiro}, {Hernandez Toledo}, {Troup}, {Trump}, {Trussler}, {Turner}, {Tuttle}, {Unda-Sanzana}, {V{\'a}zquez-Mata}, {Valentini}, {Valenzuela}, {Vargas-Gonz{\'a}lez}, {Vargas-Maga{\~n}a}, {Alfaro}, {Villanova}, {Vincenzo}, {Wake}, {Warfield}, {Washington}, {Weaver}, {Weijmans}, {Weinberg}, {Weiss}, {Westfall}, {Wild}, {Wilde}, {Wilson}, {Wilson}, {Wilson}, {Wolf}, {Wood-Vasey}, {Yan}, {Zamora}, {Zasowski}, {Zhang}, {Zhao}, {Zheng}, {Zheng}, \& {Zhu}}]{SDSSDR172022ApJS..259...35A}
{Abdurro'uf}, {Accetta}, K., {Aerts}, C., {et~al.} 2022, \apjs, 259, 35, \dodoi{10.3847/1538-4365/ac4414}

\bibitem[{{Aggarwal} {et~al.}(2021){Aggarwal}, {Budav{\'a}ri}, {Deller}, {Eftekhari}, {James}, {Prochaska}, \& {Tendulkar}}]{PATH2021ApJ...911...95A}
{Aggarwal}, K., {Budav{\'a}ri}, T., {Deller}, A.~T., {et~al.} 2021, \apj, 911, 95, \dodoi{10.3847/1538-4357/abe8d2}

\bibitem[{{Alam} {et~al.}(2015){Alam}, {Albareti}, {Allende Prieto}, {Anders}, {Anderson}, {Anderton}, {Andrews}, {Armengaud}, {Aubourg}, {Bailey}, {Basu}, {Bautista}, {Beaton}, {Beers}, {Bender}, {Berlind}, {Beutler}, {Bhardwaj}, {Bird}, {Bizyaev}, {Blake}, {Blanton}, {Blomqvist}, {Bochanski}, {Bolton}, {Bovy}, {Shelden Bradley}, {Brandt}, {Brauer}, {Brinkmann}, {Brown}, {Brownstein}, {Burden}, {Burtin}, {Busca}, {Cai}, {Capozzi}, {Carnero Rosell}, {Carr}, {Carrera}, {Chambers}, {Chaplin}, {Chen}, {Chiappini}, {Chojnowski}, {Chuang}, {Clerc}, {Comparat}, {Covey}, {Croft}, {Cuesta}, {Cunha}, {da Costa}, {Da Rio}, {Davenport}, {Dawson}, {De Lee}, {Delubac}, {Deshpande}, {Dhital}, {Dutra-Ferreira}, {Dwelly}, {Ealet}, {Ebelke}, {Edmondson}, {Eisenstein}, {Ellsworth}, {Elsworth}, {Epstein}, {Eracleous}, {Escoffier}, {Esposito}, {Evans}, {Fan}, {Fern{\'a}ndez-Alvar}, {Feuillet}, {Filiz Ak}, {Finley}, {Finoguenov}, {Flaherty}, {Fleming}, {Font-Ribera}, {Foster}, {Frinchaboy}, {Galbraith-Frew}, {Garc{\'\i}a},
  {Garc{\'\i}a-Hern{\'a}ndez}, {Garc{\'\i}a P{\'e}rez}, {Gaulme}, {Ge}, {G{\'e}nova-Santos}, {Georgakakis}, {Ghezzi}, {Gillespie}, {Girardi}, {Goddard}, {Gontcho}, {Gonz{\'a}lez Hern{\'a}ndez}, {Grebel}, {Green}, {Grieb}, {Grieves}, {Gunn}, {Guo}, {Harding}, {Hasselquist}, {Hawley}, {Hayden}, {Hearty}, {Hekker}, {Ho}, {Hogg}, {Holley-Bockelmann}, {Holtzman}, {Honscheid}, {Huber}, {Huehnerhoff}, {Ivans}, {Jiang}, {Johnson}, {Kinemuchi}, {Kirkby}, {Kitaura}, {Klaene}, {Knapp}, {Kneib}, {Koenig}, {Lam}, {Lan}, {Lang}, {Laurent}, {Le Goff}, {Leauthaud}, {Lee}, {Lee}, {Licquia}, {Liu}, {Long}, {L{\'o}pez-Corredoira}, {Lorenzo-Oliveira}, {Lucatello}, {Lundgren}, {Lupton}, {Mack}, {Mahadevan}, {Maia}, {Majewski}, {Malanushenko}, {Malanushenko}, {Manchado}, {Manera}, {Mao}, {Maraston}, {Marchwinski}, {Margala}, {Martell}, {Martig}, {Masters}, {Mathur}, {McBride}, {McGehee}, {McGreer}, {McMahon}, {M{\'e}nard}, {Menzel}, {Merloni}, {M{\'e}sz{\'a}ros}, {Miller}, {Miralda-Escud{\'e}}, {Miyatake}, {Montero-Dorta}, {More},
  {Morganson}, {Morice-Atkinson}, {Morrison}, {Mosser}, {Muna}, {Myers}, {Nandra}, {Newman}, {Neyrinck}, {Nguyen}, {Nichol}, {Nidever}, {Noterdaeme}, {Nuza}, {O'Connell}, {O'Connell}, {O'Connell}, {Ogando}, {Olmstead}, {Oravetz}, {Oravetz}, {Osumi}, {Owen}, {Padgett}, {Padmanabhan}, {Paegert}, {Palanque-Delabrouille}, {Pan}, {Parejko}, {P{\^a}ris}, {Park}, {Pattarakijwanich}, {Pellejero-Ibanez}, {Pepper}, {Percival}, {P{\'e}rez-Fournon}, {P{\'e}rez-R{\`a}fols}, {Petitjean}, {Pieri}, {Pinsonneault}, {Porto de Mello}, {Prada}, {Prakash}, {Price-Whelan}, {Protopapas}, {Raddick}, {Rahman}, {Reid}, {Rich}, {Rix}, {Robin}, {Rockosi}, {Rodrigues}, {Rodr{\'\i}guez-Torres}, {Roe}, {Ross}, {Ross}, {Rossi}, {Ruan}, {Rubi{\~n}o-Mart{\'\i}n}, {Rykoff}, {Salazar-Albornoz}, {Salvato}, {Samushia}, {S{\'a}nchez}, {Santiago}, {Sayres}, {Schiavon}, {Schlegel}, {Schmidt}, {Schneider}, {Schultheis}, {Schwope}, {Sc{\'o}ccola}, {Scott}, {Sellgren}, {Seo}, {Serenelli}, {Shane}, {Shen}, {Shetrone}, {Shu}, {Silva Aguirre}, {Sivarani},
  {Skrutskie}, {Slosar}, {Smith}, {Sobreira}, {Souto}, {Stassun}, {Steinmetz}, {Stello}, {Strauss}, {Streblyanska}, {Suzuki}, {Swanson}, {Tan}, {Tayar}, {Terrien}, {Thakar}, {Thomas}, {Thomas}, {Thompson}, {Tinker}, {Tojeiro}, {Troup}, {Vargas-Maga{\~n}a}, {Vazquez}, {Verde}, {Viel}, {Vogt}, {Wake}, {Wang}, {Weaver}, {Weinberg}, {Weiner}, {White}, {Wilson}, {Wisniewski}, {Wood-Vasey}, {Ye`che}, {York}, {Zakamska}, {Zamora}, {Zasowski}, {Zehavi}, {Zhao}, {Zheng}, {Zhou}, {Zhou}, {Zou}, \& {Zhu}}]{SDSS2015ApJS..219...12A}
{Alam}, S., {Albareti}, F.~D., {Allende Prieto}, C., {et~al.} 2015, \apjs, 219, 12, \dodoi{10.1088/0067-0049/219/1/12}

\bibitem[{{Aller}(1984)}]{Aller1984ASSL..112.....A}
{Aller}, L.~H. 1984, {Physics of thermal gaseous nebulae} (Springer Dordrecht), \dodoi{10.1007/978-94-010-9639-3}

\bibitem[{{Alloin} {et~al.}(1979){Alloin}, {Collin-Souffrin}, {Joly}, \& {Vigroux}}]{Alloin1979}
{Alloin}, D., {Collin-Souffrin}, S., {Joly}, M., \& {Vigroux}, L. 1979, \aap, 78, 200

\bibitem[{{Almeida} {et~al.}(2023){Almeida}, {Anderson}, {Argudo-Fern{\'a}ndez}, {Badenes}, {Barger}, {Barrera-Ballesteros}, {Bender}, {Benitez}, {Besser}, {Bird}, {Bizyaev}, {Blanton}, {Bochanski}, {Bovy}, {Brandt}, {Brownstein}, {Buchner}, {Bulbul}, {Burchett}, {Cano D{\'\i}az}, {Carlberg}, {Casey}, {Chandra}, {Cherinka}, {Chiappini}, {Coker}, {Comparat}, {Conroy}, {Contardo}, {Cortes}, {Covey}, {Crane}, {Cunha}, {Dabbieri}, {Davidson}, {Davis}, {de Andrade Queiroz}, {De Lee}, {M{\'e}ndez Delgado}, {Demasi}, {Di Mille}, {Donor}, {Dow}, {Dwelly}, {Eracleous}, {Eriksen}, {Fan}, {Farr}, {Frederick}, {Fries}, {Frinchaboy}, {G{\"a}nsicke}, {Ge}, {Gonz{\'a}lez {\'A}vila}, {Grabowski}, {Grier}, {Guiglion}, {Gupta}, {Hall}, {Hawkins}, {Hayes}, {Hermes}, {Hern{\'a}ndez-Garc{\'\i}a}, {Hogg}, {Holtzman}, {Ibarra-Medel}, {Ji}, {Jofre}, {Johnson}, {Jones}, {Kinemuchi}, {Kluge}, {Koekemoer}, {Kollmeier}, {Kounkel}, {Krishnarao}, {Krumpe}, {Lacerna}, {Lago}, {Laporte}, {Liu}, {Liu}, {Liu}, {Lopes}, {Macktoobian},
  {Majewski}, {Malanushenko}, {Maoz}, {Masseron}, {Masters}, {Matijevic}, {McBride}, {Medan}, {Merloni}, {Morrison}, {Myers}, {M{\'e}sz{\'a}ros}, {Negrete}, {Nidever}, {Nitschelm}, {Oravetz}, {Oravetz}, {Pan}, {Peng}, {Pinsonneault}, {Pogge}, {Qiu}, {Ramirez}, {Rix}, {Fern{\'a}ndez Rosso}, {Runnoe}, {Salvato}, {Sanchez}, {Santana}, {Saydjari}, {Sayres}, {Schlaufman}, {Schneider}, {Schwope}, {Serna}, {Shen}, {Sobeck}, {Song}, {Souto}, {Spoo}, {Stassun}, {Steinmetz}, {Straumit}, {Stringfellow}, {S{\'a}nchez-Gallego}, {Taghizadeh-Popp}, {Tayar}, {Thakar}, {Tissera}, {Tkachenko}, {Hernandez Toledo}, {Trakhtenbrot}, {Fern{\'a}ndez-Trincado}, {Troup}, {Trump}, {Tuttle}, {Ulloa}, {Vazquez-Mata}, {Vera Alfaro}, {Villanova}, {Wachter}, {Weijmans}, {Wheeler}, {Wilson}, {Wojno}, {Wolf}, {Xue}, {Ybarra}, {Zari}, \& {Zasowski}}]{2023ApJS..267...44A}
{Almeida}, A., {Anderson}, S.~F., {Argudo-Fern{\'a}ndez}, M., {et~al.} 2023, \apjs, 267, 44, \dodoi{10.3847/1538-4365/acda98}

\bibitem[{Anderson(2009)}]{MANOVAanderson2009introduction}
Anderson, T. 2009, AN INTRODUCTION TO MULTIVARIATE STATISTICAL ANALYSIS, 3RD ED (Wiley India Pvt. Limited).
\newblock \url{https://books.google.com/books?id=1iF0CgAAQBAJ}

\bibitem[{{Asplund} {et~al.}(2021){Asplund}, {Amarsi}, \& {Grevesse}}]{solar3-2021A&A...653A.141A}
{Asplund}, M., {Amarsi}, A.~M., \& {Grevesse}, N. 2021, \aap, 653, A141, \dodoi{10.1051/0004-6361/202140445}

\bibitem[{{Assef} {et~al.}(2010){Assef}, {Kochanek}, {Brodwin}, {Cool}, {Forman}, {Gonzalez}, {Hickox}, {Jones}, {Le Floc'h}, {Moustakas}, {Murray}, \& {Stern}}]{2010ApJ...713..970A}
{Assef}, R.~J., {Kochanek}, C.~S., {Brodwin}, M., {et~al.} 2010, \apj, 713, 970, \dodoi{10.1088/0004-637X/713/2/970}

\bibitem[{{Assef} {et~al.}(2015){Assef}, {Eisenhardt}, {Stern}, {Tsai}, {Wu}, {Wylezalek}, {Blain}, {Bridge}, {Donoso}, {Gonzales}, {Griffith}, \& {Jarrett}}]{2015ApJ...804...27A}
{Assef}, R.~J., {Eisenhardt}, P.~R.~M., {Stern}, D., {et~al.} 2015, \apj, 804, 27, \dodoi{10.1088/0004-637X/804/1/27}

\bibitem[{{Astropy Collaboration} {et~al.}(2022){Astropy Collaboration}, {Price-Whelan}, {Lim}, {Earl}, {Starkman}, {Bradley}, {Shupe}, {Patil}, {Corrales}, {Brasseur}, {N{\"o}the}, {Donath}, {Tollerud}, {Morris}, {Ginsburg}, {Vaher}, {Weaver}, {Tocknell}, {Jamieson}, {van Kerkwijk}, {Robitaille}, {Merry}, {Bachetti}, {G{\"u}nther}, {Aldcroft}, {Alvarado-Montes}, {Archibald}, {B{\'o}di}, {Bapat}, {Barentsen}, {Baz{\'a}n}, {Biswas}, {Boquien}, {Burke}, {Cara}, {Cara}, {Conroy}, {Conseil}, {Craig}, {Cross}, {Cruz}, {D'Eugenio}, {Dencheva}, {Devillepoix}, {Dietrich}, {Eigenbrot}, {Erben}, {Ferreira}, {Foreman-Mackey}, {Fox}, {Freij}, {Garg}, {Geda}, {Glattly}, {Gondhalekar}, {Gordon}, {Grant}, {Greenfield}, {Groener}, {Guest}, {Gurovich}, {Handberg}, {Hart}, {Hatfield-Dodds}, {Homeier}, {Hosseinzadeh}, {Jenness}, {Jones}, {Joseph}, {Kalmbach}, {Karamehmetoglu}, {Ka{\l}uszy{\'n}ski}, {Kelley}, {Kern}, {Kerzendorf}, {Koch}, {Kulumani}, {Lee}, {Ly}, {Ma}, {MacBride}, {Maljaars}, {Muna}, {Murphy}, {Norman},
  {O'Steen}, {Oman}, {Pacifici}, {Pascual}, {Pascual-Granado}, {Patil}, {Perren}, {Pickering}, {Rastogi}, {Roulston}, {Ryan}, {Rykoff}, {Sabater}, {Sakurikar}, {Salgado}, {Sanghi}, {Saunders}, {Savchenko}, {Schwardt}, {Seifert-Eckert}, {Shih}, {Jain}, {Shukla}, {Sick}, {Simpson}, {Singanamalla}, {Singer}, {Singhal}, {Sinha}, {Sip{\H{o}}cz}, {Spitler}, {Stansby}, {Streicher}, {{\v{S}}umak}, {Swinbank}, {Taranu}, {Tewary}, {Tremblay}, {de Val-Borro}, {Van Kooten}, {Vasovi{\'c}}, {Verma}, {de Miranda Cardoso}, {Williams}, {Wilson}, {Winkel}, {Wood-Vasey}, {Xue}, {Yoachim}, {Zhang}, {Zonca}, \& {Astropy Project Contributors}}]{Astropy2022ApJ...935..167A}
{Astropy Collaboration}, {Price-Whelan}, A.~M., {Lim}, P.~L., {et~al.} 2022, \apj, 935, 167, \dodoi{10.3847/1538-4357/ac7c74}

\bibitem[{{Baldwin} {et~al.}(1981){Baldwin}, {Phillips}, \& {Terlevich}}]{BPT1981PASP...93....5B}
{Baldwin}, J.~A., {Phillips}, M.~M., \& {Terlevich}, R. 1981, \pasp, 93, 5, \dodoi{10.1086/130766}

\bibitem[{Bassa {et~al.}(2017)Bassa, Tendulkar, Adams, Maddox, Bogdanov, Bower, Burke-Spolaor, Butler, Chatterjee, Cordes, Hessels, Kaspi, Law, Marcote, Paragi, Ransom, Scholz, Spitler, \& van Langevelde}]{Bassa_2017_121102HST}
Bassa, C.~G., Tendulkar, S.~P., Adams, E. A.~K., {et~al.} 2017, The Astrophysical Journal Letters, 843, L8, \dodoi{10.3847/2041-8213/aa7a0c}

\bibitem[{{Bell} {et~al.}(2003){Bell}, {McIntosh}, {Katz}, \& {Weinberg}}]{2003ApJS..149..289B}
{Bell}, E.~F., {McIntosh}, D.~H., {Katz}, N., \& {Weinberg}, M.~D. 2003, \apjs, 149, 289, \dodoi{10.1086/378847}

\bibitem[{{Bhandari} {et~al.}(2022){Bhandari}, {Gordon}, {Scott}, {Marnoch}, {Sridhar}, {Kumar}, {James}, {Qiu}, {Bannister}, {Deller}, {Eftekhari}, {Fong}, {Glowacki}, {Prochaska}, {Ryder}, {Shannon}, \& {Simha}}]{Bhandari2101172022arXiv221116790B}
{Bhandari}, S., {Gordon}, A.~C., {Scott}, D.~R., {et~al.} 2022, arXiv e-prints, arXiv:2211.16790, \dodoi{10.48550/arXiv.2211.16790}

\bibitem[{{Bhardwaj} {et~al.}(2024){Bhardwaj}, {Kirichenko}, \& {Gil de Paz}}]{2024ATel16613....1B}
{Bhardwaj}, M., {Kirichenko}, A., \& {Gil de Paz}, A. 2024, The Astronomer's Telegram, 16613, 1

\bibitem[{{Bhusare} {et~al.}(2024){Bhusare}, {Maan}, \& {Kumar}}]{uGMRT2024arXiv241213121B}
{Bhusare}, Y., {Maan}, Y., \& {Kumar}, A. 2024, arXiv e-prints, arXiv:2412.13121, \dodoi{10.48550/arXiv.2412.13121}

\bibitem[{{Bianchi} {et~al.}(2017){Bianchi}, {Shiao}, \& {Thilker}}]{2017ApJS..230...24B}
{Bianchi}, L., {Shiao}, B., \& {Thilker}, D. 2017, \apjs, 230, 24, \dodoi{10.3847/1538-4365/aa7053}

\bibitem[{{B{\"o}hringer} {et~al.}(2000){B{\"o}hringer}, {Voges}, {Huchra}, {McLean}, {Giacconi}, {Rosati}, {Burg}, {Mader}, {Schuecker}, {Simi{\c{c}}}, {Komossa}, {Reiprich}, {Retzlaff}, \& {Tr{\"u}mper}}]{NORAS2000ApJS..129..435B}
{B{\"o}hringer}, H., {Voges}, W., {Huchra}, J.~P., {et~al.} 2000, \apjs, 129, 435, \dodoi{10.1086/313427}

\bibitem[{{B{\"o}hringer} {et~al.}(2004){B{\"o}hringer}, {Schuecker}, {Guzzo}, {Collins}, {Voges}, {Cruddace}, {Ortiz-Gil}, {Chincarini}, {De Grandi}, {Edge}, {MacGillivray}, {Neumann}, {Schindler}, \& {Shaver}}]{REFLEX2004A&A...425..367B}
{B{\"o}hringer}, H., {Schuecker}, P., {Guzzo}, L., {et~al.} 2004, \aap, 425, 367, \dodoi{10.1051/0004-6361:20034484}

\bibitem[{{Bruni} {et~al.}(2024){Bruni}, {Piro}, {Yang}, {Palazzi}, {Nicastro}, {Rossi}, {Savaglio}, {Maiorano}, \& {Zhang}}]{Bruni2024arXiv241201478B}
{Bruni}, G., {Piro}, L., {Yang}, Y.~P., {et~al.} 2024, arXiv e-prints, arXiv:2412.01478, \dodoi{10.48550/arXiv.2412.01478}

\bibitem[{{Buat} {et~al.}(2005){Buat}, {Iglesias-P{\'a}ramo}, {Seibert}, {Burgarella}, {Charlot}, {Martin}, {Xu}, {Heckman}, {Boissier}, {Boselli}, {Barlow}, {Bianchi}, {Byun}, {Donas}, {Forster}, {Friedman}, {Jelinski}, {Lee}, {Madore}, {Malina}, {Milliard}, {Morissey}, {Neff}, {Rich}, {Schiminovitch}, {Siegmund}, {Small}, {Szalay}, {Welsh}, \& {Wyder}}]{2005ApJ...619L..51B}
{Buat}, V., {Iglesias-P{\'a}ramo}, J., {Seibert}, M., {et~al.} 2005, \apjl, 619, L51, \dodoi{10.1086/423241}

\bibitem[{{Bulbul} {et~al.}(2024){Bulbul}, {Liu}, {Kluge}, {Zhang}, {Sanders}, {Bahar}, {Ghirardini}, {Artis}, {Seppi}, {Garrel}, {Ramos-Ceja}, {Comparat}, {Balzer}, {B{\"o}ckmann}, {Br{\"u}ggen}, {Clerc}, {Dennerl}, {Dolag}, {Freyberg}, {Grandis}, {Gruen}, {Kleinebreil}, {Krippendorf}, {Lamer}, {Merloni}, {Migkas}, {Nandra}, {Pacaud}, {Predehl}, {Reiprich}, {Schrabback}, {Veronica}, {Weller}, \& {Zelmer}}]{eRosita2024A&A...685A.106B}
{Bulbul}, E., {Liu}, A., {Kluge}, M., {et~al.} 2024, \aap, 685, A106, \dodoi{10.1051/0004-6361/202348264}

\bibitem[{{Burgarella} {et~al.}(2005){Burgarella}, {Buat}, \& {Iglesias-P{\'a}ramo}}]{2005MNRAS.360.1413B}
{Burgarella}, D., {Buat}, V., \& {Iglesias-P{\'a}ramo}, J. 2005, \mnras, 360, 1413, \dodoi{10.1111/j.1365-2966.2005.09131.x}

\bibitem[{{Cardelli} {et~al.}(1989){Cardelli}, {Clayton}, \& {Mathis}}]{Cardelli1989ApJ...345..245C}
{Cardelli}, J.~A., {Clayton}, G.~C., \& {Mathis}, J.~S. 1989, \apj, 345, 245, \dodoi{10.1086/167900}

\bibitem[{{Chambers} {et~al.}(2016){Chambers}, {Magnier}, {Metcalfe}, {Flewelling}, {Huber}, {Waters}, {Denneau}, {Draper}, {Farrow}, {Finkbeiner}, {Holmberg}, {Koppenhoefer}, {Price}, {Rest}, {Saglia}, {Schlafly}, {Smartt}, {Sweeney}, {Wainscoat}, {Burgett}, {Chastel}, {Grav}, {Heasley}, {Hodapp}, {Jedicke}, {Kaiser}, {Kudritzki}, {Luppino}, {Lupton}, {Monet}, {Morgan}, {Onaka}, {Shiao}, {Stubbs}, {Tonry}, {White}, {Ba{\~n}ados}, {Bell}, {Bender}, {Bernard}, {Boegner}, {Boffi}, {Botticella}, {Calamida}, {Casertano}, {Chen}, {Chen}, {Cole}, {Deacon}, {Frenk}, {Fitzsimmons}, {Gezari}, {Gibbs}, {Goessl}, {Goggia}, {Gourgue}, {Goldman}, {Grant}, {Grebel}, {Hambly}, {Hasinger}, {Heavens}, {Heckman}, {Henderson}, {Henning}, {Holman}, {Hopp}, {Ip}, {Isani}, {Jackson}, {Keyes}, {Koekemoer}, {Kotak}, {Le}, {Liska}, {Long}, {Lucey}, {Liu}, {Martin}, {Masci}, {McLean}, {Mindel}, {Misra}, {Morganson}, {Murphy}, {Obaika}, {Narayan}, {Nieto-Santisteban}, {Norberg}, {Peacock}, {Pier}, {Postman}, {Primak}, {Rae}, {Rai},
  {Riess}, {Riffeser}, {Rix}, {R{\"o}ser}, {Russel}, {Rutz}, {Schilbach}, {Schultz}, {Scolnic}, {Strolger}, {Szalay}, {Seitz}, {Small}, {Smith}, {Soderblom}, {Taylor}, {Thomson}, {Taylor}, {Thakar}, {Thiel}, {Thilker}, {Unger}, {Urata}, {Valenti}, {Wagner}, {Walder}, {Walter}, {Watters}, {Werner}, {Wood-Vasey}, \& {Wyse}}]{2016arXiv161205560C}
{Chambers}, K.~C., {Magnier}, E.~A., {Metcalfe}, N., {et~al.} 2016, arXiv e-prints, arXiv:1612.05560, \dodoi{10.48550/arXiv.1612.05560}

\bibitem[{{Chatterjee} {et~al.}(2017){Chatterjee}, {Law}, {Wharton}, {Burke-Spolaor}, {Hessels}, {Bower}, {Cordes}, {Tendulkar}, {Bassa}, {Demorest}, {Butler}, {Seymour}, {Scholz}, {Abruzzo}, {Bogdanov}, {Kaspi}, {Keimpema}, {Lazio}, {Marcote}, {McLaughlin}, {Paragi}, {Ransom}, {Rupen}, {Spitler}, \& {van Langevelde}}]{Chatterjee121102-2017Natur.541...58C}
{Chatterjee}, S., {Law}, C.~J., {Wharton}, R.~S., {et~al.} 2017, \nat, 541, 58, \dodoi{10.1038/nature20797}

\bibitem[{{Cid Fernandes} {et~al.}(2010){Cid Fernandes}, {Stasi{\'n}ska}, {Schlickmann}, {Mateus}, {Vale Asari}, {Schoenell}, \& {Sodr{\'e}}}]{Fernandes2010MNRAS.403.1036C}
{Cid Fernandes}, R., {Stasi{\'n}ska}, G., {Schlickmann}, M.~S., {et~al.} 2010, \mnras, 403, 1036, \dodoi{10.1111/j.1365-2966.2009.16185.x}

\bibitem[{{Cordes} \& {Lazio}(2002)}]{NE2001-12002astro.ph..7156C}
{Cordes}, J.~M., \& {Lazio}, T.~J.~W. 2002, arXiv e-prints, astro, \dodoi{10.48550/arXiv.astro-ph/0207156}

\bibitem[{{Cordes} \& {Lazio}(2003)}]{NE2001-22003astro.ph..1598C}
---. 2003, arXiv e-prints, astro, \dodoi{10.48550/arXiv.astro-ph/0301598}

\bibitem[{{Cordes} {et~al.}(2016){Cordes}, {Wharton}, {Spitler}, {Chatterjee}, \& {Wasserman}}]{Cordes2016arXiv160505890C}
{Cordes}, J.~M., {Wharton}, R.~S., {Spitler}, L.~G., {Chatterjee}, S., \& {Wasserman}, I. 2016, arXiv e-prints, arXiv:1605.05890, \dodoi{10.48550/arXiv.1605.05890}

\bibitem[{{Dey} {et~al.}(2019){Dey}, {Schlegel}, {Lang}, {Blum}, {Burleigh}, {Fan}, {Findlay}, {Finkbeiner}, {Herrera}, {Juneau}, {Landriau}, {Levi}, {McGreer}, {Meisner}, {Myers}, {Moustakas}, {Nugent}, {Patej}, {Schlafly}, {Walker}, {Valdes}, {Weaver}, {Y{\`e}che}, {Zou}, {Zhou}, {Abareshi}, {Abbott}, {Abolfathi}, {Aguilera}, {Alam}, {Allen}, {Alvarez}, {Annis}, {Ansarinejad}, {Aubert}, {Beechert}, {Bell}, {BenZvi}, {Beutler}, {Bielby}, {Bolton}, {Brice{\~n}o}, {Buckley-Geer}, {Butler}, {Calamida}, {Carlberg}, {Carter}, {Casas}, {Castander}, {Choi}, {Comparat}, {Cukanovaite}, {Delubac}, {DeVries}, {Dey}, {Dhungana}, {Dickinson}, {Ding}, {Donaldson}, {Duan}, {Duckworth}, {Eftekharzadeh}, {Eisenstein}, {Etourneau}, {Fagrelius}, {Farihi}, {Fitzpatrick}, {Font-Ribera}, {Fulmer}, {G{\"a}nsicke}, {Gaztanaga}, {George}, {Gerdes}, {Gontcho}, {Gorgoni}, {Green}, {Guy}, {Harmer}, {Hernandez}, {Honscheid}, {Huang}, {James}, {Jannuzi}, {Jiang}, {Joyce}, {Karcher}, {Karkar}, {Kehoe}, {Kneib}, {Kueter-Young}, {Lan},
  {Lauer}, {Le Guillou}, {Le Van Suu}, {Lee}, {Lesser}, {Perreault Levasseur}, {Li}, {Mann}, {Marshall}, {Mart{\'\i}nez-V{\'a}zquez}, {Martini}, {du Mas des Bourboux}, {McManus}, {Meier}, {M{\'e}nard}, {Metcalfe}, {Mu{\~n}oz-Guti{\'e}rrez}, {Najita}, {Napier}, {Narayan}, {Newman}, {Nie}, {Nord}, {Norman}, {Olsen}, {Paat}, {Palanque-Delabrouille}, {Peng}, {Poppett}, {Poremba}, {Prakash}, {Rabinowitz}, {Raichoor}, {Rezaie}, {Robertson}, {Roe}, {Ross}, {Ross}, {Rudnick}, {Safonova}, {Saha}, {S{\'a}nchez}, {Savary}, {Schweiker}, {Scott}, {Seo}, {Shan}, {Silva}, {Slepian}, {Soto}, {Sprayberry}, {Staten}, {Stillman}, {Stupak}, {Summers}, {Sien Tie}, {Tirado}, {Vargas-Maga{\~n}a}, {Vivas}, {Wechsler}, {Williams}, {Yang}, {Yang}, {Yapici}, {Zaritsky}, {Zenteno}, {Zhang}, {Zhang}, {Zhou}, \& {Zhou}}]{LS2019AJ....157..168D}
{Dey}, A., {Schlegel}, D.~J., {Lang}, D., {et~al.} 2019, \aj, 157, 168, \dodoi{10.3847/1538-3881/ab089d}

\bibitem[{{Dong} {et~al.}(2024){Dong}, {Eftekhari}, {Fong}, {Deller}, {Mannings}, {Simha}, {Sridhar}, {Rafelski}, {Gordon}, {Bhandari}, {Day}, {Heintz}, {Hessels}, {Leja}, {James}, {Kilpatrick}, {Mahony}, {Marcote}, {Margalit}, {Nimmo}, {Prochaska}, {Escorial}, {Ryder}, {Schroeder}, {Shannon}, \& {Tejos}}]{Dong201124A2024ApJ...961...44D}
{Dong}, Y., {Eftekhari}, T., {Fong}, W.-f., {et~al.} 2024, \apj, 961, 44, \dodoi{10.3847/1538-4357/ad0cbd}

\bibitem[{{Draine}(2011)}]{DraineISM2011piim.book.....D}
{Draine}, B.~T. 2011, {Physics of the Interstellar and Intergalactic Medium} (Princeton University Press)

\bibitem[{{Eisenhardt} {et~al.}(2020){Eisenhardt}, {Marocco}, {Fowler}, {Meisner}, {Kirkpatrick}, {Garcia}, {Jarrett}, {Koontz}, {Marchese}, {Stanford}, {Caselden}, {Cushing}, {Cutri}, {Faherty}, {Gelino}, {Gonzalez}, {Mainzer}, {Mobasher}, {Schlegel}, {Stern}, {Teplitz}, \& {Wright}}]{2020ApJS..247...69E}
{Eisenhardt}, P. R.~M., {Marocco}, F., {Fowler}, J.~W., {et~al.} 2020, \apjs, 247, 69, \dodoi{10.3847/1538-4365/ab7f2a}

\bibitem[{{Feng} {et~al.}(2022){Feng}, {Li}, {Yang}, {Zhang}, {Zhu}, {Zhang}, {Lu}, {Wang}, {Dai}, {Lynch}, {Yao}, {Jiang}, {Niu}, {Zhou}, {Xu}, {Miao}, {Niu}, {Meng}, {Qian}, {Tsai}, {Wang}, {Xue}, {Yue}, {Yuan}, {Zhang}, \& {Zhang}}]{FengRM2022Sci...375.1266F}
{Feng}, Y., {Li}, D., {Yang}, Y.-P., {et~al.} 2022, Science, 375, 1266, \dodoi{10.1126/science.abl7759}

\bibitem[{{Fitzpatrick} \& {Massa}(2007)}]{2007ApJ...663..320F}
{Fitzpatrick}, E.~L., \& {Massa}, D. 2007, \apj, 663, 320, \dodoi{10.1086/518158}

\bibitem[{{Fong} {et~al.}(2021){Fong}, Dong, Leja, Bhandari, Day, Deller, Kumar, Prochaska, Scott, Bannister, Eftekhari, Gordon, Heintz, James, Kilpatrick, Mahony, Escorial, Ryder, Shannon, \& Tejos}]{201124AFong_2021}
{Fong}, Dong, Y., Leja, J., {et~al.} 2021, The Astrophysical Journal Letters, 919, L23, \dodoi{10.3847/2041-8213/ac242b}

\bibitem[{{Foreman-Mackey} {et~al.}(2013){Foreman-Mackey}, {Hogg}, {Lang}, \& {Goodman}}]{emcmc2013PASP..125..306F}
{Foreman-Mackey}, D., {Hogg}, D.~W., {Lang}, D., \& {Goodman}, J. 2013, \pasp, 125, 306, \dodoi{10.1086/670067}

\bibitem[{{Fujita} \& {Aung}(2019)}]{clusterhalo2019ApJ...875...26F}
{Fujita}, Y., \& {Aung}, H. 2019, \apj, 875, 26, \dodoi{10.3847/1538-4357/ab0e02}

\bibitem[{{Gordon} {et~al.}(2023){Gordon}, {Fong}, {Kilpatrick}, {Eftekhari}, {Leja}, {Prochaska}, {Nugent}, {Bhandari}, {Blanchard}, {Caleb}, {Day}, {Deller}, {Dong}, {Glowacki}, {Gourdji}, {Mannings}, {Mahoney}, {Marnoch}, {Miller}, {Paterson}, {Rastinejad}, {Ryder}, {Sadler}, {Scott}, {Sears}, {Shannon}, {Simha}, {Stappers}, \& {Tejos}}]{Gordon2023hostgalaxies}
{Gordon}, A.~C., {Fong}, W.-f., {Kilpatrick}, C.~D., {et~al.} 2023, \apj, 954, 80, \dodoi{10.3847/1538-4357/ace5aa}

\bibitem[{{Gordon} {et~al.}(2003){Gordon}, {Clayton}, {Misselt}, {Landolt}, \& {Wolff}}]{2003ApJ...594..279G}
{Gordon}, K.~D., {Clayton}, G.~C., {Misselt}, K.~A., {Landolt}, A.~U., \& {Wolff}, M.~J. 2003, \apj, 594, 279, \dodoi{10.1086/376774}

\bibitem[{{Graham} \& {Fruchter}(2017)}]{Graham2017ApJ...834..170G}
{Graham}, J.~F., \& {Fruchter}, A.~S. 2017, \apj, 834, 170, \dodoi{10.3847/1538-4357/834/2/170}

\bibitem[{{Griffith} {et~al.}(2011){Griffith}, {Tsai}, {Stern}, {Blain}, {Eisenhardt}, {Harrison}, {Jarrett}, {Madsen}, {Stanford}, {Wright}, {Wu}, {Wu}, \& {Yan}}]{Griffith2011ApJ...736L..22G}
{Griffith}, R.~L., {Tsai}, C.-W., {Stern}, D., {et~al.} 2011, \apjl, 736, L22, \dodoi{10.1088/2041-8205/736/1/L22}

\bibitem[{{Haffner} {et~al.}(2009){Haffner}, {Dettmar}, {Beckman}, {Wood}, {Slavin}, {Giammanco}, {Madsen}, {Zurita}, \& {Reynolds}}]{WIM2009RvMP...81..969H}
{Haffner}, L.~M., {Dettmar}, R.~J., {Beckman}, J.~E., {et~al.} 2009, Reviews of Modern Physics, 81, 969, \dodoi{10.1103/RevModPhys.81.969}

\bibitem[{{Hidalgo-G{\'a}mez}(2007)}]{Hidalgo-G´amez2007}
{Hidalgo-G{\'a}mez}, A.~M. 2007, \aj, 134, 1447, \dodoi{10.1086/521392}

\bibitem[{{Ibik} {et~al.}(2024){Ibik}, {Drout}, {Gaensler}, {Scholz}, {Sridhar}, {Margalit}, {Clarke}, {Tendulkar}, {Michilli}, {Eftekhari}, {Bhardwaj}, {Burke-Spolaor}, {Chatterjee}, {Cook}, {Hessels}, {Kirsten}, {Joseph}, {Kaspi}, {Lazda}, {Masui}, {Nimmo}, {Pandhi}, {Pearlman}, {Pleunis}, {Rafiei-Ravandi}, {Shin}, \& {Smith}}]{newPRS2024arXiv240911533I}
{Ibik}, A.~L., {Drout}, M.~R., {Gaensler}, B.~M., {et~al.} 2024, arXiv e-prints, arXiv:2409.11533.
\newblock \doarXiv{2409.11533}

\bibitem[{{Izotov} {et~al.}(2006){Izotov}, {Stasi{\'n}ska}, {Meynet}, {Guseva}, \& {Thuan}}]{Izotov2006A&A...448..955I}
{Izotov}, Y.~I., {Stasi{\'n}ska}, G., {Meynet}, G., {Guseva}, N.~G., \& {Thuan}, T.~X. 2006, \aap, 448, 955, \dodoi{10.1051/0004-6361:20053763}

\bibitem[{{Kauffmann} {et~al.}(2003{\natexlab{a}}){Kauffmann}, {Heckman}, {White}, {Charlot}, {Tremonti}, {Brinchmann}, {Bruzual}, {Peng}, {Seibert}, {Bernardi}, {Blanton}, {Brinkmann}, {Castander}, {Cs{\'a}bai}, {Fukugita}, {Ivezic}, {Munn}, {Nichol}, {Padmanabhan}, {Thakar}, {Weinberg}, \& {York}}]{Kauffmann2003MNRAS.341...33K}
{Kauffmann}, G., {Heckman}, T.~M., {White}, S. D.~M., {et~al.} 2003{\natexlab{a}}, \mnras, 341, 33, \dodoi{10.1046/j.1365-8711.2003.06291.x}

\bibitem[{{Kauffmann} {et~al.}(2003{\natexlab{b}}){Kauffmann}, {Heckman}, {Tremonti}, {Brinchmann}, {Charlot}, {White}, {Ridgway}, {Brinkmann}, {Fukugita}, {Hall}, {Ivezi{\'c}}, {Richards}, \& {Schneider}}]{Kauffmann2003MNRAS.346.1055K}
{Kauffmann}, G., {Heckman}, T.~M., {Tremonti}, C., {et~al.} 2003{\natexlab{b}}, \mnras, 346, 1055, \dodoi{10.1111/j.1365-2966.2003.07154.x}

\bibitem[{{Kennicutt}(1998)}]{1998ARA&A..36..189K}
{Kennicutt}, Robert~C., J. 1998, \araa, 36, 189, \dodoi{10.1146/annurev.astro.36.1.189}

\bibitem[{{Kennicutt} \& {Evans}(2012)}]{2012ARA&A..50..531K}
{Kennicutt}, R.~C., \& {Evans}, N.~J. 2012, \araa, 50, 531, \dodoi{10.1146/annurev-astro-081811-125610}

\bibitem[{{Kewley} {et~al.}(2006){Kewley}, {Groves}, {Kauffmann}, \& {Heckman}}]{Kewley2006MNRAS.372..961K}
{Kewley}, L.~J., {Groves}, B., {Kauffmann}, G., \& {Heckman}, T. 2006, \mnras, 372, 961, \dodoi{10.1111/j.1365-2966.2006.10859.x}

\bibitem[{{Kobulnicky} \& {Kewley}(2004)}]{KK042004ApJ...617..240K}
{Kobulnicky}, H.~A., \& {Kewley}, L.~J. 2004, \apj, 617, 240, \dodoi{10.1086/425299}

\bibitem[{{Kokubo} {et~al.}(2017){Kokubo}, {Mitsuda}, {Sugai}, {Ozaki}, {Minowa}, {Hattori}, {Hayano}, {Matsubayashi}, {Shimono}, {Sako}, \& {Doi}}]{Kokubo_2017_121102IFU}
{Kokubo}, M., {Mitsuda}, K., {Sugai}, H., {et~al.} 2017, \apj, 844, 95, \dodoi{10.3847/1538-4357/aa7b2d}

\bibitem[{{Kumar} {et~al.}(2024){Kumar}, {Maan}, \& {Bhusare}}]{2024ATel16452....1K}
{Kumar}, A., {Maan}, Y., \& {Bhusare}, Y. 2024, The Astronomer's Telegram, 16452, 1

\bibitem[{{Law} {et~al.}(2022){Law}, {Connor}, \& {Aggarwal}}]{PRS2022ApJ...927...55L}
{Law}, C.~J., {Connor}, L., \& {Aggarwal}, K. 2022, \apj, 927, 55, \dodoi{10.3847/1538-4357/ac4c42}

\bibitem[{{Lee} {et~al.}(2023){Lee}, {Khrykin}, {Simha}, {Ata}, {Huang}, {Prochaska}, {Tejos}, {Cooke}, {Nagamine}, \& {Zhang}}]{190520foreground2023ApJ...954L...7L}
{Lee}, K.-G., {Khrykin}, I.~S., {Simha}, S., {et~al.} 2023, \apjl, 954, L7, \dodoi{10.3847/2041-8213/acefb5}

\bibitem[{{Li} {et~al.}(2023){Li}, {Tsai}, {Stern}, {Wu}, {Assef}, {Blain}, {D{\'\i}az-Santos}, {Eisenhardt}, {Griffith}, {Jarrett}, {Jun}, {Lake}, \& {Saade}}]{2023ApJ...958..162L}
{Li}, G., {Tsai}, C.-W., {Stern}, D., {et~al.} 2023, \apj, 958, 162, \dodoi{10.3847/1538-4357/ace25b}

\bibitem[{{Macquart} {et~al.}(2020){Macquart}, {Prochaska}, {McQuinn}, {Bannister}, {Bhandari}, {Day}, {Deller}, {Ekers}, {James}, {Marnoch}, {Os{\l}owski}, {Phillips}, {Ryder}, {Scott}, {Shannon}, \& {Tejos}}]{MacquartRelation2020Natur.581..391M}
{Macquart}, J.~P., {Prochaska}, J.~X., {McQuinn}, M., {et~al.} 2020, \nat, 581, 391, \dodoi{10.1038/s41586-020-2300-2}

\bibitem[{{Marocco} {et~al.}(2021){Marocco}, {Eisenhardt}, {Fowler}, {Kirkpatrick}, {Meisner}, {Schlafly}, {Stanford}, {Garcia}, {Caselden}, {Cushing}, {Cutri}, {Faherty}, {Gelino}, {Gonzalez}, {Jarrett}, {Koontz}, {Mainzer}, {Marchese}, {Mobasher}, {Schlegel}, {Stern}, {Teplitz}, \& {Wright}}]{2021ApJS..253....8M}
{Marocco}, F., {Eisenhardt}, P. R.~M., {Fowler}, J.~W., {et~al.} 2021, \apjs, 253, 8, \dodoi{10.3847/1538-4365/abd805}

\bibitem[{{Martin}(1997)}]{Martin1997}
{Martin}, C.~L. 1997, \apj, 491, 561, \dodoi{10.1086/304978}

\bibitem[{{Niu} {et~al.}(2022){Niu}, {Aggarwal}, {Li}, {Zhang}, {Chatterjee}, {Tsai}, {Yu}, {Law}, {Burke-Spolaor}, {Cordes}, {Zhang}, {Ocker}, {Yao}, {Wang}, {Feng}, {Niino}, {Bochenek}, {Cruces}, {Connor}, {Jiang}, {Dai}, {Luo}, {Li}, {Miao}, {Niu}, {Anna-Thomas}, {Sydnor}, {Stern}, {Wang}, {Yuan}, {Yue}, {Zhou}, {Yan}, {Zhu}, \& {Zhang}}]{Niu190520-2022Natur.606..873N}
{Niu}, C.~H., {Aggarwal}, K., {Li}, D., {et~al.} 2022, \nat, 606, 873, \dodoi{10.1038/s41586-022-04755-5}

\bibitem[{{Ocker} {et~al.}(2022){Ocker}, {Cordes}, {Chatterjee}, {Niu}, {Li}, {McKee}, {Law}, {Tsai}, {Anna-Thomas}, {Yao}, \& {Cruces}}]{OckerHost2022ApJ...931...87O}
{Ocker}, S.~K., {Cordes}, J.~M., {Chatterjee}, S., {et~al.} 2022, \apj, 931, 87, \dodoi{10.3847/1538-4357/ac6504}

\bibitem[{{O'Connor} {et~al.}(2024){O'Connor}, {Bhardwaj}, \& {Palmese}}]{2024ATel16426....1O}
{O'Connor}, B., {Bhardwaj}, M., \& {Palmese}, A. 2024, The Astronomer's Telegram, 16426, 1

\bibitem[{{Ould-Boukattine} {et~al.}(2024){Ould-Boukattine}, {Hessels}, {Kirsten}, {Hewitt}, {Snelders}, {Blaauw}, {Sluman}, {Mulder}, {Herrmann}, {Gawronski}, {Puchalska}, \& {Gopinath}}]{2024ATel16432....1O}
{Ould-Boukattine}, O.~S., {Hessels}, J.~W.~T., {Kirsten}, F., {et~al.} 2024, The Astronomer's Telegram, 16432, 1

\bibitem[{{Panda} {et~al.}(2024){Panda}, {Bhattacharyya}, {Dudeja}, {Kudale}, \& {Roy}}]{2024ATel16494....1P}
{Panda}, U., {Bhattacharyya}, S., {Dudeja}, C., {Kudale}, S., \& {Roy}, J. 2024, The Astronomer's Telegram, 16494, 1

\bibitem[{{Pelliciari} {et~al.}(2024){Pelliciari}, {Geminardi}, {Bernardi}, {Pilia}, {Esposito}, \& {Naldi}}]{2024ATel16434....1P}
{Pelliciari}, D., {Geminardi}, A., {Bernardi}, G., {et~al.} 2024, The Astronomer's Telegram, 16434, 1

\bibitem[{{Pettini} \& {Pagel}(2004)}]{PP042004MNRAS.348L..59P}
{Pettini}, M., \& {Pagel}, B. E.~J. 2004, \mnras, 348, L59, \dodoi{10.1111/j.1365-2966.2004.07591.x}

\bibitem[{{Piffaretti} {et~al.}(2011){Piffaretti}, {Arnaud}, {Pratt}, {Pointecouteau}, \& {Melin}}]{MCXC2011A&A...534A.109P}
{Piffaretti}, R., {Arnaud}, M., {Pratt}, G.~W., {Pointecouteau}, E., \& {Melin}, J.~B. 2011, \aap, 534, A109, \dodoi{10.1051/0004-6361/201015377}

\bibitem[{{Piro} {et~al.}(2021){Piro}, {Bruni, G.}, {Troja, E.}, {O'Connor, B.}, {Panessa, F.}, {Ricci, R.}, {Zhang, B.}, {Burgay, M.}, {Dichiara, S.}, {Lee, K. J.}, {Lotti, S.}, {Niu, J. R.}, {Pilia, M.}, {Possenti, A.}, {Trudu, M.}, {Xu, H.}, {Zhu, W. W.}, {Kutyrev, A. S.}, \& {Veilleux, S.}}]{201124APiro2021}
{Piro}, {Bruni, G.}, {Troja, E.}, {et~al.} 2021, A\&A, 656, L15, \dodoi{10.1051/0004-6361/202141903}

\bibitem[{{Planck Collaboration} {et~al.}(2020){Planck Collaboration}, {Aghanim}, {Akrami}, {Ashdown}, {Aumont}, {Baccigalupi}, {Ballardini}, {Banday}, {Barreiro}, {Bartolo}, {Basak}, {Battye}, {Benabed}, {Bernard}, {Bersanelli}, {Bielewicz}, {Bock}, {Bond}, {Borrill}, {Bouchet}, {Boulanger}, {Bucher}, {Burigana}, {Butler}, {Calabrese}, {Cardoso}, {Carron}, {Challinor}, {Chiang}, {Chluba}, {Colombo}, {Combet}, {Contreras}, {Crill}, {Cuttaia}, {de Bernardis}, {de Zotti}, {Delabrouille}, {Delouis}, {Di Valentino}, {Diego}, {Dor{\'e}}, {Douspis}, {Ducout}, {Dupac}, {Dusini}, {Efstathiou}, {Elsner}, {En{\ss}lin}, {Eriksen}, {Fantaye}, {Farhang}, {Fergusson}, {Fernandez-Cobos}, {Finelli}, {Forastieri}, {Frailis}, {Fraisse}, {Franceschi}, {Frolov}, {Galeotta}, {Galli}, {Ganga}, {G{\'e}nova-Santos}, {Gerbino}, {Ghosh}, {Gonz{\'a}lez-Nuevo}, {G{\'o}rski}, {Gratton}, {Gruppuso}, {Gudmundsson}, {Hamann}, {Handley}, {Hansen}, {Herranz}, {Hildebrandt}, {Hivon}, {Huang}, {Jaffe}, {Jones}, {Karakci}, {Keih{\"a}nen},
  {Keskitalo}, {Kiiveri}, {Kim}, {Kisner}, {Knox}, {Krachmalnicoff}, {Kunz}, {Kurki-Suonio}, {Lagache}, {Lamarre}, {Lasenby}, {Lattanzi}, {Lawrence}, {Le Jeune}, {Lemos}, {Lesgourgues}, {Levrier}, {Lewis}, {Liguori}, {Lilje}, {Lilley}, {Lindholm}, {L{\'o}pez-Caniego}, {Lubin}, {Ma}, {Mac{\'\i}as-P{\'e}rez}, {Maggio}, {Maino}, {Mandolesi}, {Mangilli}, {Marcos-Caballero}, {Maris}, {Martin}, {Martinelli}, {Mart{\'\i}nez-Gonz{\'a}lez}, {Matarrese}, {Mauri}, {McEwen}, {Meinhold}, {Melchiorri}, {Mennella}, {Migliaccio}, {Millea}, {Mitra}, {Miville-Desch{\^e}nes}, {Molinari}, {Montier}, {Morgante}, {Moss}, {Natoli}, {N{\o}rgaard-Nielsen}, {Pagano}, {Paoletti}, {Partridge}, {Patanchon}, {Peiris}, {Perrotta}, {Pettorino}, {Piacentini}, {Polastri}, {Polenta}, {Puget}, {Rachen}, {Reinecke}, {Remazeilles}, {Renzi}, {Rocha}, {Rosset}, {Roudier}, {Rubi{\~n}o-Mart{\'\i}n}, {Ruiz-Granados}, {Salvati}, {Sandri}, {Savelainen}, {Scott}, {Shellard}, {Sirignano}, {Sirri}, {Spencer}, {Sunyaev}, {Suur-Uski}, {Tauber}, {Tavagnacco},
  {Tenti}, {Toffolatti}, {Tomasi}, {Trombetti}, {Valenziano}, {Valiviita}, {Van Tent}, {Vibert}, {Vielva}, {Villa}, {Vittorio}, {Wandelt}, {Wehus}, {White}, {White}, {Zacchei}, \& {Zonca}}]{Planck2020A&A...641A...6P}
{Planck Collaboration}, {Aghanim}, N., {Akrami}, Y., {et~al.} 2020, \aap, 641, A6, \dodoi{10.1051/0004-6361/201833910}

\bibitem[{{Pleunis} {et~al.}(2021){Pleunis}, {Good}, {Kaspi}, {Mckinven}, {Ransom}, {Scholz}, {Bandura}, {Bhardwaj}, {Boyle}, {Brar}, {Cassanelli}, {Chawla}, {(Adam) Dong}, {Fonseca}, {Gaensler}, {Josephy}, {Kaczmarek}, {Leung}, {Lin}, {Masui}, {Mena-Parra}, {Michilli}, {Ng}, {Patel}, {Rafiei-Ravandi}, {Rahman}, {Sanghavi}, {Shin}, {Smith}, {Stairs}, \& {Tendulkar}}]{Pleunis2021ApJ...923....1P}
{Pleunis}, Z., {Good}, D.~C., {Kaspi}, V.~M., {et~al.} 2021, \apj, 923, 1, \dodoi{10.3847/1538-4357/ac33ac}

\bibitem[{{Polles} {et~al.}(2019){Polles}, {Madden}, {Lebouteiller}, {Cormier}, {Abel}, {Galliano}, {Hony}, {Karczewski}, {Lee}, {Chevance}, {Galametz}, \& {Lianou}}]{Polles2019}
{Polles}, F.~L., {Madden}, S.~C., {Lebouteiller}, V., {et~al.} 2019, \aap, 622, A119, \dodoi{10.1051/0004-6361/201833776}

\bibitem[{{Popesso} {et~al.}(2023){Popesso}, {Concas}, {Cresci}, {Belli}, {Rodighiero}, {Inami}, {Dickinson}, {Ilbert}, {Pannella}, \& {Elbaz}}]{SFMSredshift2023MNRAS.519.1526P}
{Popesso}, P., {Concas}, A., {Cresci}, G., {et~al.} 2023, \mnras, 519, 1526, \dodoi{10.1093/mnras/stac3214}

\bibitem[{{Press} \& {Davis}(1982)}]{Press1982}
{Press}, W.~H., \& {Davis}, M. 1982, \apj, 259, 449, \dodoi{10.1086/160183}

\bibitem[{{Prochaska} {et~al.}(2020{\natexlab{a}}){Prochaska}, Hennawi, Westfall, Cooke, Wang, Hsyu, Davies, Farina, \& Pelliccia}]{pypeit1}
{Prochaska}, J.~X., Hennawi, J.~F., Westfall, K.~B., {et~al.} 2020{\natexlab{a}}, Journal of Open Source Software, 5, 2308, \dodoi{10.21105/joss.02308}

\bibitem[{{Prochaska} {et~al.}(2020{\natexlab{b}}){Prochaska}, {Hennawi}, {Cooke}, {Westfall}, {Wang}, {EmAstro}, {Tiffanyhsyu}, {Wasserman}, {Villaume}, {Marijana777}, {Schindler}, {Young}, {Simha}, {Wilde}, {Tejos}, {Isbell}, {Fl{\"o}rs}, {Sandford}, {Vasovi{\'c}}, {Betts}, \& {Holden}}]{pypeit2}
{Prochaska}, J.~X., {Hennawi}, J., {Cooke}, R., {et~al.} 2020{\natexlab{b}}, {pypeit/PypeIt: Release 1.0.0}, v1.0.0,  Zenodo, \dodoi{10.5281/zenodo.3743493}

\bibitem[{{Reynolds}(1977)}]{1977ApJ...216..433R}
{Reynolds}, R.~J. 1977, \apj, 216, 433, \dodoi{10.1086/155484}

\bibitem[{{Sadat} {et~al.}(2004){Sadat}, {Blanchard}, {Kneib}, {Mathez}, {Madore}, \& {Mazzarella}}]{Sadat2004A&A...424.1097S}
{Sadat}, R., {Blanchard}, A., {Kneib}, J.~P., {et~al.} 2004, \aap, 424, 1097, \dodoi{10.1051/0004-6361:20034029}

\bibitem[{{Salpeter}(1955)}]{1955ApJ...121..161S}
{Salpeter}, E.~E. 1955, \apj, 121, 161, \dodoi{10.1086/145971}

\bibitem[{{Schlafly} \& {Finkbeiner}(2011)}]{2011ApJ...737..103S}
{Schlafly}, E.~F., \& {Finkbeiner}, D.~P. 2011, \apj, 737, 103, \dodoi{10.1088/0004-637X/737/2/103}

\bibitem[{{Schlegel} {et~al.}(1998){Schlegel}, {Finkbeiner}, \& {Davis}}]{1998ApJ...500..525S}
{Schlegel}, D.~J., {Finkbeiner}, D.~P., \& {Davis}, M. 1998, \apj, 500, 525, \dodoi{10.1086/305772}

\bibitem[{{Sharma} {et~al.}(2024){Sharma}, {Ravi}, {Connor}, {Law}, {Ocker}, {Sherman}, {Kosogorov}, {Faber}, {Hallinan}, {Harnach}, {Hellbourg}, {Hobbs}, {Hodge}, {Hodges}, {Lamb}, {Rasmussen}, {Somalwar}, {Weinreb}, {Woody}, {Leja}, {Anand}, {Kashyap Das}, {Qin}, {Rose}, {Dong}, {Miller}, \& {Yao}}]{Sharma2024arXiv240916964S}
{Sharma}, K., {Ravi}, V., {Connor}, L., {et~al.} 2024, arXiv e-prints, arXiv:2409.16964, \dodoi{10.48550/arXiv.2409.16964}

\bibitem[{{Shin} \& {CHIME/FRB Collaboration}(2024)}]{CHIMEdiscovery2024ATel16420....1S}
{Shin}, K., \& {CHIME/FRB Collaboration}. 2024, The Astronomer's Telegram, 16420, 1

\bibitem[{{Snelders} {et~al.}(2024){Snelders}, {Bhandari}, {Kirsten}, {Hessels}, {Marcote}, {Hewitt}, {Gawronski}, {Puchalska}, {Ould-Boukattine}, {Gopinath}, {Nimmo}, {Karuppusamy}, {Herrmann}, {Yang}, {Blaauw}, {Buttaccio}, {Maccaferri}, {Bach}, {Feiler}, {Bray}, {Williams}, {Wrigley}, {Keimpema}, {Paragi}, {Burgay}, {Corongiu}, {Giroletti}, {Kramer}, {Pilia}, {Spitler}, {Surcis}, {Trudu}, {Yuan}, {Wang}, \& {Bezrukovs}}]{2024ATel16542....1S}
{Snelders}, M.~P., {Bhandari}, S., {Kirsten}, F., {et~al.} 2024, The Astronomer's Telegram, 16542, 1

\bibitem[{{Spitler} {et~al.}(2014){Spitler}, {Cordes}, {Hessels}, {Lorimer}, {McLaughlin}, {Chatterjee}, {Crawford}, {Deneva}, {Kaspi}, {Wharton}, {Allen}, {Bogdanov}, {Brazier}, {Camilo}, {Freire}, {Jenet}, {Karako-Argaman}, {Knispel}, {Lazarus}, {Lee}, {van Leeuwen}, {Lynch}, {Ransom}, {Scholz}, {Siemens}, {Stairs}, {Stovall}, {Swiggum}, {Venkataraman}, {Zhu}, {Aulbert}, \& {Fehrmann}}]{Spitler2014ApJ...790..101S}
{Spitler}, L.~G., {Cordes}, J.~M., {Hessels}, J.~W.~T., {et~al.} 2014, \apj, 790, 101, \dodoi{10.1088/0004-637X/790/2/101}

\bibitem[{{Spitler} {et~al.}(2016){Spitler}, {Scholz}, {Hessels}, {Bogdanov}, {Brazier}, {Camilo}, {Chatterjee}, {Cordes}, {Crawford}, {Deneva}, {Ferdman}, {Freire}, {Kaspi}, {Lazarus}, {Lynch}, {Madsen}, {McLaughlin}, {Patel}, {Ransom}, {Seymour}, {Stairs}, {Stappers}, {van Leeuwen}, \& {Zhu}}]{Spitler2016Natur.531..202S}
{Spitler}, L.~G., {Scholz}, P., {Hessels}, J.~W.~T., {et~al.} 2016, \nat, 531, 202, \dodoi{10.1038/nature17168}

\bibitem[{{Tempel} {et~al.}(2012){Tempel}, {Tago}, \& {Liivam{\"a}gi}}]{tempel2012}
{Tempel}, E., {Tago}, E., \& {Liivam{\"a}gi}, L.~J. 2012, \aap, 540, A106, \dodoi{10.1051/0004-6361/201118687}

\bibitem[{{Tempel} {et~al.}(2014){Tempel}, {Tamm}, {Gramann}, {Tuvikene}, {Liivam{\"a}gi}, {Suhhonenko}, {Kipper}, {Einasto}, \& {Saar}}]{tempel2014}
{Tempel}, E., {Tamm}, A., {Gramann}, M., {et~al.} 2014, \aap, 566, A1, \dodoi{10.1051/0004-6361/201423585}

\bibitem[{{Tendulkar} {et~al.}(2017){Tendulkar}, {Bassa}, {Cordes}, {Bower}, {Law}, {Chatterjee}, {Adams}, {Bogdanov}, {Burke-Spolaor}, {Butler}, {Demorest}, {Hessels}, {Kaspi}, {Lazio}, {Maddox}, {Marcote}, {McLaughlin}, {Paragi}, {Ransom}, {Scholz}, {Seymour}, {Spitler}, {van Langevelde}, \& {Wharton}}]{Tendulkar121102-2017ApJ...834L...7T}
{Tendulkar}, S.~P., {Bassa}, C.~G., {Cordes}, J.~M., {et~al.} 2017, \apjl, 834, L7, \dodoi{10.3847/2041-8213/834/2/L7}

\bibitem[{{Tian} {et~al.}(2024{\natexlab{a}}){Tian}, {Pastor-Marazuela}, {Stappers}, {Rajwade}, {Caleb}, {Bezuidenhout}, {Barr}, \& {Kramer}}]{2024ATel16446....1T}
{Tian}, J., {Pastor-Marazuela}, I., {Stappers}, B., {et~al.} 2024{\natexlab{a}}, The Astronomer's Telegram, 16446, 1

\bibitem[{{Tian} {et~al.}(2024{\natexlab{b}}){Tian}, {Rajwade}, {Pastor-Marazuela}, {Stappers}, {Bezuidenhout}, {Caleb}, {Jankowski}, {Barr}, \& {Kramer}}]{Tian2024MNRAS.533.3174T}
{Tian}, J., {Rajwade}, K.~M., {Pastor-Marazuela}, I., {et~al.} 2024{\natexlab{b}}, \mnras, 533, 3174, \dodoi{10.1093/mnras/stae2013}

\bibitem[{{Turner} \& {Gott}(1976)}]{Turner1976}
{Turner}, E.~L., \& {Gott}, J.~R., I. 1976, \apjs, 32, 409, \dodoi{10.1086/190403}

\bibitem[{{Uttarkar} {et~al.}(2024){Uttarkar}, {Kumar}, {Lower}, \& {Shannon}}]{2024ATel16430....1U}
{Uttarkar}, P.~A., {Kumar}, P., {Lower}, M.~E., \& {Shannon}, R.~M. 2024, The Astronomer's Telegram, 16430, 1

\bibitem[{Virtanen {et~al.}(2020)Virtanen, Gommers, Oliphant, Haberland, Reddy, Cournapeau, Burovski, Peterson, Weckesser, Bright, {van der Walt}, Brett, Wilson, Millman, Mayorov, Nelson, Jones, Kern, Larson, Carey, Polat, Feng, Moore, {VanderPlas}, Laxalde, Perktold, Cimrman, Henriksen, Quintero, Harris, Archibald, Ribeiro, Pedregosa, {van Mulbregt}, \& {SciPy 1.0 Contributors}}]{2020SciPy-NMeth}
Virtanen, P., Gommers, R., Oliphant, T.~E., {et~al.} 2020, Nature Methods, 17, 261, \dodoi{10.1038/s41592-019-0686-2}

\bibitem[{{Wen} \& {Han}(2024)}]{Wen2024ApJS..272...39W}
{Wen}, Z.~L., \& {Han}, J.~L. 2024, \apjs, 272, 39, \dodoi{10.3847/1538-4365/ad409d}

\bibitem[{{Wright} {et~al.}(2010){Wright}, {Eisenhardt}, {Mainzer}, {Ressler}, {Cutri}, {Jarrett}, {Kirkpatrick}, {Padgett}, {McMillan}, {Skrutskie}, {Stanford}, {Cohen}, {Walker}, {Mather}, {Leisawitz}, {Gautier}, {McLean}, {Benford}, {Lonsdale}, {Blain}, {Mendez}, {Irace}, {Duval}, {Liu}, {Royer}, {Heinrichsen}, {Howard}, {Shannon}, {Kendall}, {Walsh}, {Larsen}, {Cardon}, {Schick}, {Schwalm}, {Abid}, {Fabinsky}, {Naes}, \& {Tsai}}]{2010AJ....140.1868W}
{Wright}, E.~L., {Eisenhardt}, P. R.~M., {Mainzer}, A.~K., {et~al.} 2010, \aj, 140, 1868, \dodoi{10.1088/0004-6256/140/6/1868}

\bibitem[{{Yamasaki} \& {Totani}(2020)}]{DMhalo2020ApJ...888..105Y}
{Yamasaki}, S., \& {Totani}, T. 2020, \apj, 888, 105, \dodoi{10.3847/1538-4357/ab58c4}

\bibitem[{{Yao} {et~al.}(2017){Yao}, {Manchester}, \& {Wang}}]{YMW2017ApJ...835...29Y}
{Yao}, J.~M., {Manchester}, R.~N., \& {Wang}, N. 2017, \apj, 835, 29, \dodoi{10.3847/1538-4357/835/1/29}

\bibitem[{{Zhang}(2018)}]{ZhangDMz2018ApJ...867L..21Z}
{Zhang}, B. 2018, \apjl, 867, L21, \dodoi{10.3847/2041-8213/aae8e3}

\bibitem[{{Zhang} {et~al.}(2024{\natexlab{a}}){Zhang}, {Zhu}, {Cao}, {Zhou}, {Zhang}, {Xie}, {Wu}, {Wang}, {Wang}, {Niu}, {Di Li}, {Zhu}, {Zhang}, {Han}, {Lee}, {Wang}, {Gao}, {Feng}, {Jiang}, {Jing}, {Li}, {Lu}, {Luo}, {Lyu}, {Wang}, {Xu}, {Yang}, {Yu}, {Zhang}, \& {Project}}]{2024ATel16433....1Z}
{Zhang}, J., {Zhu}, Y., {Cao}, S., {et~al.} 2024{\natexlab{a}}, The Astronomer's Telegram, 16433, 1

\bibitem[{{Zhang} {et~al.}(2024{\natexlab{b}}){Zhang}, {Wu}, {Cao}, {Zhu}, {Zhang}, {Niu}, {Xie}, {Zhou}, {Wang}, {Zhu}, {Zhang}, {Wang}, {Niu}, {Di Li}, {Han}, {Lee}, {Wang}, {Gao}, {Feng}, {Jiang}, {Jing}, {Li}, {Lu}, {Luo}, {Lyu}, {Wang}, {Xu}, {Yang}, {Yu}, {Zhang}, \& {Project}}]{2024ATel16505....1Z}
{Zhang}, J., {Wu}, Q., {Cao}, S., {et~al.} 2024{\natexlab{b}}, The Astronomer's Telegram, 16505, 1

\bibitem[{{Zhang} \& {Yu}(2024)}]{2024ATel16695....1Z}
{Zhang}, X., \& {Yu}, W. 2024, The Astronomer's Telegram, 16695, 1

\bibitem[{{Zhang} {et~al.}(2025){Zhang}, {Yu}, {Yan}, {Xing}, \& {Zhang}}]{PRSYu2025arXiv250114247Z}
{Zhang}, X., {Yu}, W., {Yan}, Z., {Xing}, Y., \& {Zhang}, B. 2025, arXiv e-prints, arXiv:2501.14247, \dodoi{10.48550/arXiv.2501.14247}

\end{thebibliography}
\bibliographystyle{aasjournal}

\end{document}